\documentclass[%
reprint,
amsmath,amssymb,
aps,
]{revtex4-2}
\usepackage{subcaption}
\usepackage{graphicx}
\usepackage{dcolumn}
\usepackage{bm}
\usepackage{physics}
\usepackage{amsmath}
\usepackage{color}
\begin{document}
	\title{Practical Phase-Coding Side-Channel-Secure Quantum Key Distribution}
	\author{Yang-Guang Shan}
	\author{Zhen-Qiang Yin}
	\email{yinzq@ustc.edu.cn}
	\author{Shuang Wang}
	\email{wshuang@ustc.edu.cn}
	\author{Wei Chen}
	\author{De-Yong He}
	\author{Guang-Can Guo}
	\author{Zheng-Fu Han}
	\affiliation{CAS Key Laboratory of Quantum Information, University of Science and Technology of China, Hefei, Anhui 230026, China}
	\affiliation{CAS Center for Excellence in Quantum Information and Quantum Physics, University of Science and Technology of China, Hefei, Anhui 230026, China}
	\affiliation{State Key Laboratory of Cryptology, P. O. Box 5159, Beijing 100878, China}
	
	\begin{abstract}
	All kinds of device loopholes give rise to a great obstacle to practical secure quantum key distribution (QKD). In this article, inspired by the original side-channel-secure protocol [Physical Review Applied 12, 054034 (2019)], a new QKD protocol called phase-coding side-channel-secure (PC-SCS) protocol is proposed. This protocol can be immune to all uncorrelated side channels of the source part and all loopholes of the measurement side. A finite-key security analysis against coherent attack of the new protocol is given. The proposed protocol only requires modulation of two phases, which can avoid the challenge of preparing perfect vacuum states. Numerical simulation shows that a practical transmission distance of $300$ km can be realized by the PC-SCS protocol.
	\end{abstract}
\maketitle
	\section{introduction}
	Quantum key distribution (QKD) \cite{bennett1984quantum,renner2008security} promises to realize unconditional secure key distribution between two distant peers (usually called Alice and Bob). It is also one of the most practical technologies in quantum information science. Though the theory of QKD seems to be unassailable, practical devices may not always meet the requirement of protocols. Thus an eavesdropper, Eve, may steal information without the discovery of Alice and Bob. All kinds of loopholes of devices \cite{brassard2000limitations,lutkenhaus2002quantum,qi2007time,inamori2007unconditional,wang2008general,PhysRevA.78.042333,scarani2009security,makarov2009controlling,lydersen2010hacking,wiechers2011after,lu2023hacking} established a great obstacle to realizing a practically secure QKD system. 
	
	The ultimate solution to all loopholes must be the device-independent (DI) QKD \cite{mayers1998quantum,acin2007device}, which promises to be secure without characterizing any operating principles of devices. However, DI QKD is hard to realize and its performance on transmission distance and key rate is quite bad. Thus some trade-off between security and performance must be conducted in practical use. Luckily, the measurement-device-independent (MDI) QKD protocol \cite{lo2012measurement,braunstein2012side} can be immune to all loopholes of the measurement part, which is the most vulnerable part in QKD systems, and high performance can also be realized at the same time. Based on MDI, twin-field (TF) QKD \cite{lucamarini2018overcoming,ma2018phase,wang2018twin,lin2018simple,curty2019simple,cui2019twin} and mode-pairing QKD \cite{zeng2022mode} (see also in \cite{xie2022breaking}) can even break the repeaterless secret-key capacity bounds \cite{pirandola2009direct,takeoka2014fundamental,pirandola2017fundamental} with both high security and performance. In addition, the measurement part of MDI can be fully controlled by an eavesdropper, and the measurement results can be known by an eavesdropper, which cannot be leaked in DI QKD.
	
	 In recent years, QKD protocols considering side channels of the source side \cite{tamaki2014loss,pereira2019quantum,wang2019practical,pereira2020quantum,navarrete2021practical,gu2022experimental} have attracted some attention because of the requirement of higher security than the MDI QKD. In \cite{tamaki2014loss,pereira2019quantum,pereira2020quantum,navarrete2021practical,gu2022experimental}, the authors developed an ingenious method called reference technique to deal with the flaws of the source side. However, this method requires a detailed characterization of all possible side channels to analyze their influence, which seems to be a difficult task in practice. 
	 
	 In \cite{wang2019practical}, a new protocol called side-channel-secure (SCS) QKD is proposed based on the sending-or-not-sending (SNS) TF QKD \cite{wang2018twin}. In this protocol, the users do not need to care about the details of the source side channels. It only requires that the upper bound of the pulse intensity is known to ensure security. An experiment \cite{zhang2022experimental} has shown its practicality in 50-km fibre and a practical SCS QKD with more than 100-km transmission distance can be wished to come true. However, vacuum states are needed in the original SCS (SNS-SCS) protocol, and a small intensity of imperfect vacuum states can drastically influence its performance \cite{jiang2022side}. Another problem is that the security analysis of the SNS-SCS protocol is conducted under the assumption of collective attacks, and the analysis under coherent attacks is still missing. To solve this problem, we propose a phase-coding side-channel-secure (PC-SCS) protocol. In our scheme, Alice and Bob do not need to modulate different intensities. They only modulate different phases to encode information, which is easy to realize. And we can avoid the difficulty of preparing vacuum states. Our PC-SCS protocol has the same level of security as the SNS-SCS protocol, which means only the upper bound of signal intensity is trusted. We also give a finite-key analysis for our protocol under coherent attacks, which is also the first finite-key analysis for the SCS protocol. Numerical simulation shows that with a reasonable pulse number of $10^{13}$ level, our PC-SCS protocol could realize a higher key rate and a longer distance than the SNS-SCS protocol.
	 
	 \section{protocol description}
	 
	 The modulation of our protocol is quite similar to the signal states of the no-phase-postselection TF QKD \cite{lin2018simple,curty2019simple,cui2019twin}, but phase randomization and decoy states \cite{hwang2003quantum,PhysRevLett.94.230504,wang2005beating,ma2005practical} are not needed, and the users only need to modulate two phases. A schematic figure is shown in Fig. \ref{fig:setup}, and a detailed procedure is shown below.
	 \begin{figure}
	 	\centering
	 	\includegraphics[width=0.5\textwidth]{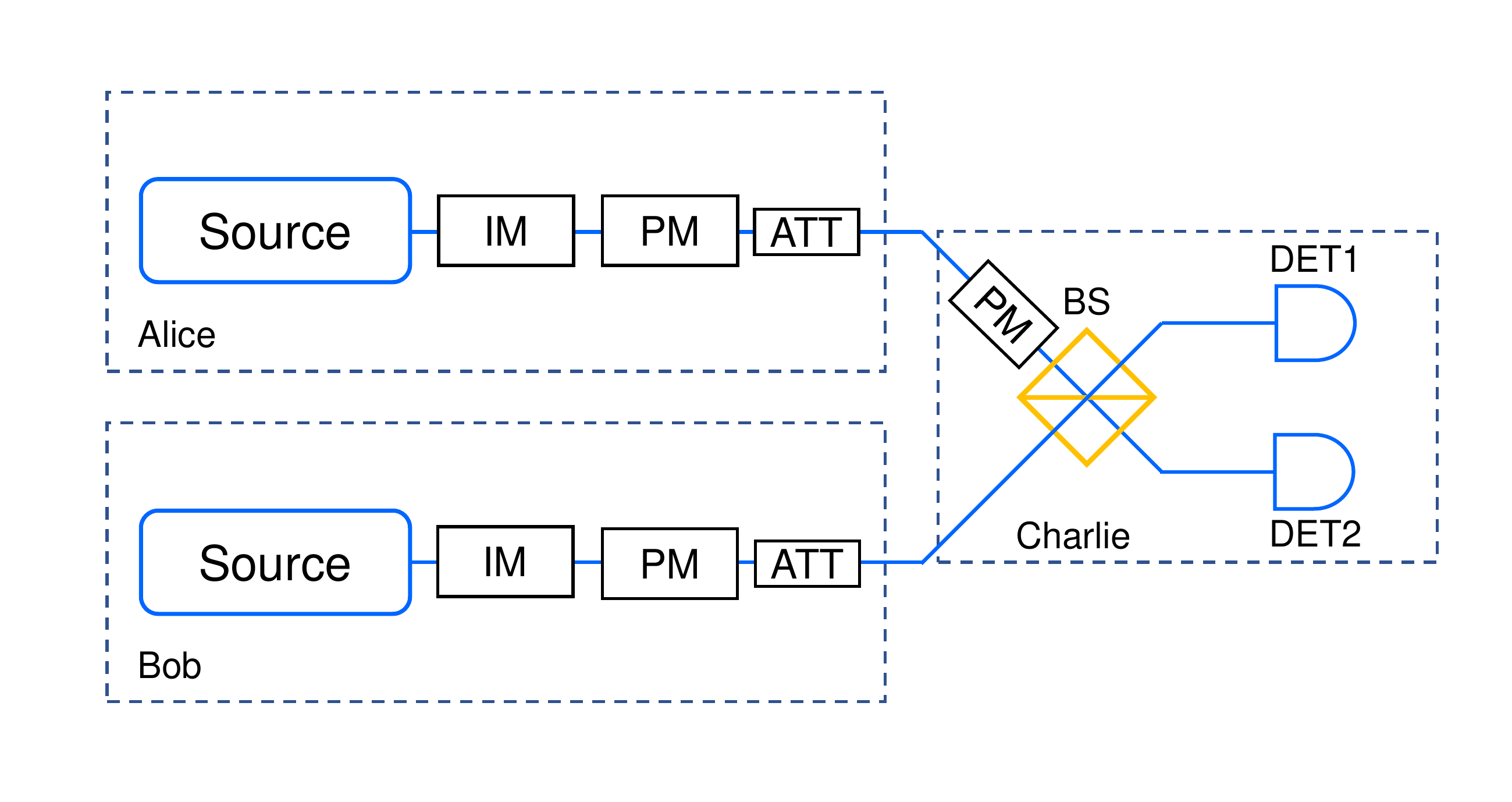}
	 	\caption{\label{fig:setup}The schematic figure of our PC-SCS protocol. IM: intensity modulator, PM: phase modulator, ATT: attenuator, BS: beam splitter, DET: detector.}
	 \end{figure}
	 \begin{enumerate}
	 	\item 
	 	 \textbf{State preparation.} With phase-locking technology, Alice and Bob prepare coherent state pulses with the same phase. The intensities of every pulse are the same. Then for every time window, Alice (Bob) chooses to modulate a phase $0$ or $\pi$ uniformly at random. When she (he) chooses to modulate phase $0$, she (he) records a classical bit $0$ locally. And when she (he) decides to modulate phase $\pi$, she (he) records a classical bit $1$. Then they coincidentally send the encoded coherent states to the untrusted peer Charlie who is located in the middle of the channel.
	 	\item
	 	 \textbf{State measurement.} If Charlie is honest, he will conduct an interference measurement for the pulses from Alice and Bob. We assume that when the phase modulations of Alice and Bob are the same, an ideal interferometer will produce a click on the left detector (DET1). And if the phase difference of the two pulses is $\pi$, an ideal interferometer will produce a click on the right detector (DET2). Note that Alice and Bob could send strong reference light to Charlie to estimate the phase shift from the channel or other devices, thus Charlie could compensate this phase shift with a phase modulator to realize this kind of measurement. If only one detector clicks, Charlie will announce a successful measurement. Charlie also announces that it is a left-click event or a right-click event. Note that if Charlie is not honest, he may conduct an arbitrary measurement on all pulses sent by Alice and Bob, and fabricate the measurement result for all time windows at one time.
	 	 \item 
	 	 \textbf{Sifting.} The first step is repeated to accumulate enough data. For the time windows Charlie announced a successful measurement, Alice and Bob will retain the corresponding classical bits as raw key bits. If Charlie announced a right-click event, Bob will flip his corresponding classical bit.
	 	\item 
	 	 \textbf{Parameter estimation.} From all time windows, Alice and Bob randomly choose a part to announce their classical bits to analyze the phase error rate. We denote that for every time window, there is a probability of $P_{est}$ to be chosen. Then the raw key bits from the rest time windows are used to form the final key bits.
	 	 \item 
	 	 \textbf{Postprocessing}. Alice and Bob conduct error correction and private amplification to the rest raw key bits to get the final secure key bits.
	 \end{enumerate}
	 \section{side channel analysis}
	 In this section, we will give a detailed explanation of the side channels we considered. Firstly, note that our protocol is not source-device-independent, which is also stressed in \cite{wang2019practical}. We assume that the eavesdropper Eve cannot hack into the source part to directly steal the encoding information. Then the correlations between time windows are not considered, however different side channels of different time windows are allowed in our analysis.
	 
	 According to our protocol description, an ideal source will randomly choose to send the two coherent states,
	 \begin{eqnarray}
	 \ket{\alpha}=\sum_{n=0}^{\infty}\text{e}^{-\frac{\abs{\alpha}^2}{2}}\frac{\alpha^n}{\sqrt{n!}}\ket{n},\\
	 \ket{-\alpha}=\sum_{n=0}^{\infty}\text{e}^{-\frac{\abs{\alpha}^2}{2}}\frac{(-\alpha)^n}{\sqrt{n!}}\ket{n},
	 \end{eqnarray}
 where $\ket{n}$ is the Fock state of $n$ photons. We note that the vacuum state $\ket{0}$ cannot be encoded by side channel information. Thus a practical source will produce states in the following form,
 \begin{eqnarray}
 	\ket{\alpha'}=\sqrt{P_{0}^+}\ket{0}+\sqrt{1-P_0^+}\ket{\varphi},\label{eq:alpha}\\
 	\ket{-\alpha'}=\sqrt{P_0^-}\ket{0}+\sqrt{1-P_0^-}\ket{\psi},\label{eq:-alpha}
 \end{eqnarray}
where $\ket{\varphi}$ and $\ket{\psi}$ are two states including the dimensions of side channels. By the theorem from \cite{wang2019practical}, we only need to know the lower bound of the amplitude of vacuum states, which is $P_0^{+},P_0^{-}\ge \underline{P}_0$. For coherent states, $\underline{P}_0$ corresponds to $\text{e}^{-\abs{\alpha}^2}$. But we do not require the state to be a coherent state. Our protocol has no restriction on the states $\ket{\varphi}$ and $\ket{\psi}$.

To describe the general side channels, in the analysis below $P_0^+$, $P_0^-$, $\ket{\varphi}$ and $\ket{\psi}$ can be different in different time windows and can be different for Alice and Bob.

\section{security analysis}
Our security analysis is based on the calculation of the so-called phase error rate of an equivalent protocol based on entanglement, which is given below.

Alice (Bob) prepares an ancilla locally entangled with the state sent out, the overall state is shown as,
	\begin{equation}
	\ket{\phi}=\frac{1}{2}\left(\ket{0}_\text{a}\ket{\alpha'}_\text{A}+\ket{1}_\text{a}\ket{-\alpha'}_\text{A}\right)\left(\ket{0}_\text{b}\ket{\alpha'}_\text{B}+\ket{1}_\text{b}\ket{-\alpha'}_\text{B}\right),
	\end{equation}
where the state with subscript ``a'' (``b'') is the local ancilla prepared by Alice (Bob), and the state with subscript ``A'' (``B'') is the state sent by Alice (Bob) to Charlie. Note that the state $\ket{\alpha'}_\text{A}$ ($\ket{-\alpha'}_\text{A}$) and $\ket{\alpha'}_\text{B}$ ($\ket{-\alpha'}_\text{B}$) can be different because of different side channels of Alice and Bob.

Then after the announcement of measurement results by Charlie, Alice and Bob will measure their ancillary qubits a and b on the $\mathbb{Z}$ basis ($\ket{0}$, $\ket{1}$) to get their classical encoding bits and then conduct the same postprocessing procedure as the original protocol to get the final key bits.

To get the phase errors, which are the errors when Alice and Bob measure the ancillary bits on the $\mathbb{X}$ basis ($\ket{+}=\frac{1}{\sqrt{2}}(\ket{0}+\ket{1})$, $\ket{-}=\frac{1}{\sqrt{2}}(\ket{0}-\ket{1})$, we rewrite the state $\ket{\phi}$ in the following.

\begin{widetext}
\begin{equation}
\begin{aligned}
\ket{\phi}=&\frac{1}{4}\ket{++}_\text{a,b}(\ket{\alpha',\alpha'}+\ket{-\alpha',-\alpha'}+\ket{\alpha',-\alpha'}+\ket{-\alpha',\alpha'})_{\text{A,B}}+\frac{1}{4}\ket{--}_\text{a,b}(\ket{\alpha',\alpha'}+\ket{-\alpha',-\alpha'}-\ket{\alpha',-\alpha'}-\ket{-\alpha',\alpha'})_{\text{A,B}}\\
&+\frac{1}{4}\ket{+-}_\text{a,b}(\ket{\alpha',\alpha'}-\ket{-\alpha',-\alpha'}-\ket{\alpha',-\alpha'}+\ket{-\alpha',\alpha'})_{\text{A,B}}+\frac{1}{4}\ket{-+}_\text{a,b}(\ket{\alpha',\alpha'}-\ket{-\alpha',-\alpha'}+\ket{\alpha',-\alpha'}-\ket{-\alpha',\alpha'})_{\text{A,B}}\\
\equiv&\ket{++}_\text{a,b}\sqrt{P_{ee}}\ket{\varphi_{ee}}_{\text{A,B}}+\ket{--}_\text{a,b}\sqrt{P_{oo}}\ket{\varphi_{oo}}_{\text{A,B}}+\ket{+-}_\text{a,b}\sqrt{P_{eo}}\ket{\varphi_{eo}}_{\text{A,B}}+\ket{-+}_\text{a,b}\sqrt{P_{oe}}\ket{\varphi_{oe}}_{\text{A,B}}
\end{aligned}\label{eq:phi0}
\end{equation}
\end{widetext}

In Eq.(\ref{eq:phi0}), we use the denotation that $\sqrt{P_{ee}}\ket{\varphi_{ee}}_{\text{A,B}}=(\ket{\alpha',\alpha'}+\ket{-\alpha',-\alpha'}+\ket{\alpha',-\alpha'}+\ket{-\alpha',\alpha'})_{\text{A,B}}/4$ because this state corresponds to the situation that photon numbers of Alice and Bob are both even when there are no side channels. For the same reason, we use $\varphi_{oo}$ for the odd-odd situation, $\varphi_{eo}$ for the even-odd situation, and $\varphi_{oe}$ for the odd-even situation.

Since each time window has a probability of $P_{est}$ to be chosen as a parameter estimation window, we assume Alice and Bob produce another ancilla with orthogonal states $\ket{sig}$ and $\ket{est}$ to express this process. When this ancilla is projected to $\ket{sig}$, the corresponding time window is used to generate the final key. When this ancilla is projected to $\ket{est}$, the classical bits of Alice and Bob are published to calculate the bit error rate, which will then be used to calculate the phase error rate. Thus the state of a single time window is expressed as $\ket{\Phi}=\sqrt{1-P_{est}}\ket{sig}\ket{\phi}+\sqrt{P_{est}}\ket{est}\ket{\phi}$.

For a protocol with $N$ time windows, the side channels and intensities of different time windows might be different, but the intensity upper bound is the same. Since the correlation between time windows is not considered, the overall state can be expressed as $\ket{\Phi}_1\otimes\ket{\Phi}_2\otimes\dots\ket{\Phi}_N$.

If there is no side channel, we can realize that the items $\ket{\varphi_{ee}}_{\text{A,B}}$, $\ket{\varphi_{oo}}_{\text{A,B}}$ from Eq.(\ref{eq:phi0}) correspond to the traveling states containing even total photons and the items $\ket{\varphi_{eo}}_{\text{A,B}}$, $\ket{\varphi_{oe}}_{\text{A,B}}$ correspond to the traveling states containing odd total photons. Note that clicks are mainly caused by single-photon states belonging to the odd-photon state if Charlie is honest. Thus we set the states $\ket{++}_{\text{a,b}}$ and $\ket{--}_{\text{a,b}}$, which correspond to even-photon items, to be phase errors. 

For ease of understanding, we will firstly introduce our security analysis under the collective attack for infinite key length. The detailed finite-key security analysis can be seen in our Appendix \ref{sec:appsec}. For simplicity, we will ignore the parameter estimation windows in our collective-attack analysis.

The general collective attack of Eve can be seen as follows. Eve chooses a set of complete measurement operators $M_e^L$, $M_e^R$ and $M_e^O$, satisfying $M_e^{L\dagger}M_e^L+M_e^{R\dagger}M_e^R+M_e^{O\dagger}M_e^O=\mathbb{I}$. $\mathbb{I}$ is the identity operator. For each time window, Eve will conduct this measurement on the two pulses sent by Alice and Bob. If the measurement result corresponds to the operator $M_e^L$ ($M_e^R$), Eve will announce that it is a successful measurement of a left-click (right-click) event. And the operator $M_e^O$ corresponds to a failed measurement. 

Due to the symmetry of our protocol, the analyses of left-click events and right-click events are the same. We will take the left-click events as an example. The probability of finding a left-click phase error can be expressed as 
\begin{equation}
	\begin{aligned}
	P_{ph}^L=&\left\Vert M_e^L\bra{++}_{\text{a,b}} \ket{\phi}\right\Vert^2+\left\Vert M_e^L\bra{--}_{\text{a,b}} \ket{\phi}\right\Vert^2\\
	=&P_{ee}\left\Vert M_e^L\ket{\varphi_{ee}}_{\text{A,B}}\right\Vert^2+P_{oo}\left\Vert M_e^L\ket{\varphi_{oo}}_{\text{A,B}}\right\Vert^2,
	\end{aligned}\label{eq:phc}
\end{equation}
which is the probability that Eve announces a left-click measurement and Alice and Bob get $\ket{++}$ or $\ket{--}$ by measuring their ancillas. 

With the same method, we can also calculate the probability of finding a left-click bit error, which is 
\begin{equation}
	\begin{aligned}
		P_{err}^L=&\frac{1}{2}\left\Vert M_e^L(\bra{++}_{\text{a,b}}-\bra{--}_{\text{a,b}}) \ket{\phi}\right\Vert^2\\
		&+\frac{1}{2}\left\Vert M_e^L(\bra{+-}_{\text{a,b}}-\bra{-+}_{\text{a,b}}) \ket{\phi}\right\Vert^2\\
		\ge& \frac{1}{2}\left\Vert M_e^L(\sqrt{P_{ee}}\ket{\varphi_{ee}}_{\text{A,B}}-\sqrt{P_{oo}}\ket{\varphi_{oo}}_{\text{A,B}})\right\Vert^2.
	\end{aligned}
\end{equation}
Here a bit error corresponds to $\ket{01}_{\text{a,b}}$ and $\ket{10}_{\text{a,b}}$ for the ancillas of Alice and Bob. We can also calculate it with $(\ket{01}_{\text{a,b}}+\ket{10}_{\text{a,b}})/\sqrt{2}$ and $(\ket{01}_{\text{a,b}}-\ket{10}_{\text{a,b}})/\sqrt{2}$, which correspond to $(\ket{++}_{\text{a,b}}-\ket{--}_{\text{a,b}})/\sqrt{2}$ and $(\ket{+-}_{\text{a,b}}-\ket{-+}_{\text{a,b}})/\sqrt{2}$. The inequality only keeps the first item of $(\ket{++}_{\text{a,b}}-\ket{--}_{\text{a,b}})/\sqrt{2}$.

With the triangle inequality, we can get that 
\begin{equation}
	\sqrt{2P_{err}^L}\ge\sqrt{P_{ee}}\left\Vert M_e^L\ket{\varphi_{ee}}_{\text{A,B}}\right\Vert-\sqrt{P_{oo}}\left\Vert M_e^L\ket{\varphi_{oo}}_{\text{A,B}}\right\Vert.\label{eq:bitc}
\end{equation}

With Eq. (\ref{eq:phc}, \ref{eq:bitc}), we can get the upper bound of the left-click phase error rate, which is 
\begin{equation}
	\begin{aligned}
	P_{ph}^L\le& 2P_{err}^L+2\sqrt{2}\sqrt{P_{err}^LP_{oo}}\left\Vert M_e^L\ket{\varphi_{oo}}_{\text{A,B}}\right\Vert\\
	&+2P_{oo}\left\Vert M_e^L\ket{\varphi_{oo}}_{\text{A,B}}\right\Vert^2.\\
	\end{aligned}
\end{equation}

With a same process, we can also get the upper bound of the right-click phase error rate, which is 
\begin{equation}
	\begin{aligned}
		P_{ph}^R\le& 2P_{err}^R+2\sqrt{2}\sqrt{P_{err}^RP_{oo}}\left\Vert M_e^R\ket{\varphi_{oo}}_{\text{A,B}}\right\Vert\\
		&+2P_{oo}\left\Vert M_e^R\ket{\varphi_{oo}}_{\text{A,B}}\right\Vert^2.
	\end{aligned}
\end{equation}

Then the upper bound of the total phase error rate is shown as,
\begin{equation}
	\small
	\begin{aligned}
	&P_{ph}=P_{ph}^L+P_{ph}^R\\
	\le& 2P_{err}+2\sqrt{2}\sqrt{P_{err}P_{oo}}\sqrt{\left\Vert M_e^L\ket{\varphi_{oo}}_{\text{A,B}}\right\Vert^2+\left\Vert M_e^R\ket{\varphi_{oo}}_{\text{A,B}}\right\Vert^2}\\
	&+2P_{oo}(\left\Vert M_e^L\ket{\varphi_{oo}}_{\text{A,B}}\right\Vert^2+\left\Vert M_e^R\ket{\varphi_{oo}}_{\text{A,B}}\right\Vert^2)\\
	\le& 2P_{err}+2\sqrt{2}\sqrt{P_{err}P_{oo}}+2P_{oo},
\end{aligned}
\end{equation}
where we used the Cauchy-Schwarz inequality for the first inequality. And here we have used the inequality that $\left\Vert M_e^L\ket{\varphi_{oo}}_{\text{A,B}}\right\Vert^2+\left\Vert M_e^R\ket{\varphi_{oo}}_{\text{A,B}}\right\Vert^2\le 1$, because that $\sum_{i=L,R,O}\left\Vert M_e^i\ket{\varphi_{oo}}_{\text{A,B}}\right\Vert^2=1$ for a complete set of measurement. $P_{err}=P_{err}^L+P_{err}^R$ is the total bit error rate of left-click and right-click events. The bit error rate can be measured and $P_{oo}$ can be bounded by $P_{oo}\le(1-\underline{P}_0)^2$ (see Appendix \ref{sec:appsec}). Thus we have given the upper bound of the phase error rate.

For finite-key analysis considering coherent attack, we can get a similar expression of phase errors, which is shown in Eq. (\ref{eq:nphu}). Here $N$ is the total pulse number sent by Alice (Bob). $N_{est,bit}$ is the number of bit errors from parameter estimation windows. $\mu$ is the intensity upper bound of the coherent state pulses. $\text{U}^{\epsilon^2}_{e}(\cdot)$ is the upper bound of random variables' mathematical expectation estimated with measurement result by Kato's inequality \cite{kato2020concentration}. $\text{U}^{\epsilon^2}_{m}(\cdot)$ is the upper bound of random variables' measurement result estimated with mathematical expectation by Kato's inequality. $\text{C}_{U}^{\epsilon^2}(\cdot)$ is the upper bound of random variables' measurement result estimated with mathematical expectation by Chernoff bound \cite{mitzenmacher2017probability}. $\epsilon^2$ is the corresponding failure probability. And the total failure probability of Eq. (\ref{eq:nphu}) is $4\epsilon^2$. Details of these bounds can be seen in Appendix \ref{bound}.

\begin{widetext}
\begin{equation}
	\begin{aligned}
\bar{N}_{ph}=\text{U}^{\epsilon^2}_{m}\bigg\{2\frac{1-P_{est}}{P_{est}}\text{U}^{\epsilon^2}_{e}(N_{est,bit})
+\frac{2\sqrt{2}\sqrt{(1-P_{est})}}{\sqrt{P_{est}}}\sqrt{\text{U}^{\epsilon^2}_{e}(N_{est,bit})\text{U}^{\epsilon^2}_{e}\left[\text{C}_{U}^{\epsilon^2}\left(N(1-P_{est})(1-\text{e}^{-\mu})^2\right)\right]}
\\+2\text{U}^{\epsilon^2}_{e}\left[\text{C}_{U}^{\epsilon^2}(N(1-P_{est})(1-\text{e}^{-\mu})^2)\right]\bigg\}
\end{aligned}\label{eq:nphu}
\end{equation}
\end{widetext}

To meet the composability of security \cite{muller2009composability}, the key length $l$ obeys $\epsilon_{sec}=2\epsilon'+\frac{1}{2}\sqrt{2^{l-H_{min}^{\epsilon'}(A|E')}}=2\epsilon'+\tilde{\epsilon}$ \cite{tomamichel2012tight}. $\epsilon_{sec}$ is the failure parameter of security. Thus $l=H_{min}^{\epsilon'}(A|E')-2\log_2\frac{1}{2\tilde{\epsilon}}$. Here $\epsilon'=\sqrt{4\epsilon^2}=2\epsilon$. Then the final key length becomes \cite{renner2008security}
\begin{equation}
	l=N_{sig}(1-H_2(\frac{\bar{N}_{ph}}{N_{sig}})-fH_2(e_\text{bit}))-2\log_2\frac{1}{2\tilde{\epsilon}}-\log_2\frac{2}{\epsilon_{cor}}
\end{equation}
The total secure parameter is $\epsilon_{tot}=4\epsilon+\tilde{\epsilon}+\epsilon_{cor}$.

\section{numerical simulation}
We conduct a numerical simulation to see the performance of our protocol. Here we set both $\tilde{\epsilon}$ and the correction parameter $\epsilon_{cor}$ to be $\epsilon$. Then the total key rate is
\begin{equation}
R=\frac{N_{sig}}{N}(1-H_2(\frac{\bar{N}_{ph}}{N_{sig}})-fH_2(e_\text{bit}))-\frac{1}{N}\log_2\frac{1}{2\epsilon^3},
\end{equation}
\begin{equation}
\epsilon_{tot}=6\epsilon.
\end{equation}

We define that the click rate of the detector with constructive (destructive) interference is $S_{\text{large}}$ ($S_{\text{small}}$), which could be given by
\begin{equation}
\begin{aligned}
S_{\text{large}}=(1-(1-d)\text{e}^{-(1+V)\eta\mu})(1-d)\text{e}^{-(1-V)\eta\mu},\\
S_{\text{small}}=(1-(1-d)\text{e}^{-(1-V)\eta\mu})(1-d)\text{e}^{-(1+V)\eta\mu},
\end{aligned}
\end{equation}
where $V$ is the interference visibility, whose relation with the misalignment rate $\text{e}_{\text{mis}}$ is $V=1-2\text{e}_{\text{mis}}$. $d$ is the dark counting rate. $\eta$ is the transmitting efficiency from Alice (Bob) to Charlie. And $\eta=10^{-\text{loss}/20}$, where loss is the total transmission loss (dB) from Alice to Bob.

Then the counting rate of signal windows is 
\begin{equation}
\frac{N_{sig}}{N}=(1-P_{est})(S_{\text{large}}+S_{\text{small}}).
\end{equation}
And the bit error rate can be easily got by 
\begin{equation}
\text{e}_{\text{bit}}=\frac{S_{\text{small}}}{S_{\text{large}}+S_{\text{small}}},
\end{equation}
\begin{equation}
\frac{N_{est,bit}}{N}=P_{est}(S_{\text{large}}+S_{\text{small}})\text{e}_{\text{bit}}=P_{est}S_{\text{small}}.
\end{equation}
The phase error item can be gotten from Eq. (\ref{eq:nphu}).

The simulation parameters used are shown in Table. \ref{table:pa}. $P_d$ is the detecting efficiency and $d$ is the dark counting rate per window of detectors. $f$ is the efficiency of error correction. $e_{\text{mis}}$ is the misalignment rate. And $\epsilon_{tot}$ is the total security parameter.
\begin{table}[h]
\caption{\label{table:pa}Parameters we used in our simulation.}
\begin{ruledtabular}
\begin{tabular}{ccccc}
	$P_d$&$d$&$f$&$e_\text{mis}$&$\epsilon_{tot}$\\\hline
	$0.3$&$5\times10^{-11}$&$1.1$&$0.015$&$10^{-10}$
\end{tabular}
\end{ruledtabular}
\end{table}
\begin{figure}[h]
	\centering
	\includegraphics[width=0.5\textwidth]{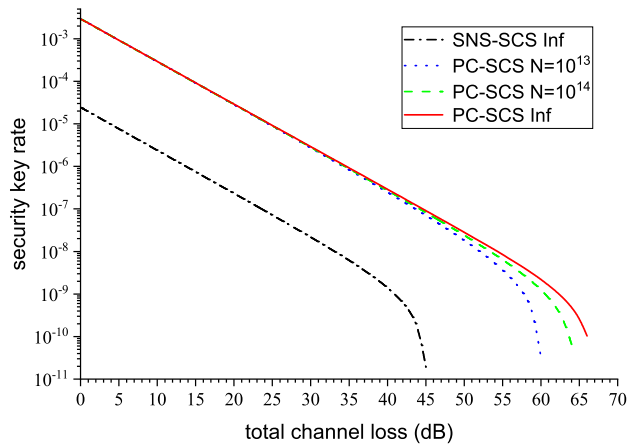}
	\caption{\label{fig:rate}The key rate comparison between our protocol and the original SNS-SCS protocol. Our PC-SCS protocol is simulated under different sent pulse number of $10^{13}$, $10^{14}$ and infinity. The SNS-SCS is simulated under the condition of infinite sent pulses.}
\end{figure}

To compare with previous work, we also simulated the SNS-SCS protocol for infinite key length. Our simulation can be seen in Fig. \ref{fig:rate}. Even with a practical finite-key length, our protocol could have a much better performance than the SNS-SCS protocol. With our protocol, a practical long-distance side-channel-secure key distribution with a 300 km transmission distance in fibre can be wished to come true.

Though our protocol seems to have a better performance, the SNS-SCS protocol has another advantage, which is its high misalignment tolerance. In \cite{wang2019practical}, the authors find that the SNS-SCS protocol can tolerate a misalignment rate of more than $30\%$, while a misalignment of about $8\%$ could prevent all key generation in our PC-SCS protocol.

\section{Conclusion}
In conclusion, we proposed a new QKD protocol called PC-SCS protocol. Our protocol could be immune to side channels of the source side and all loopholes of the measurement side, which is the same as the SNS-SCS protocol \cite{wang2019practical}. We give a complete finite-key security analysis for our protocol. With a reasonable pulse number of $10^{13}$, $10^{14}$, the performance is close to the asymptotic case. And a transmission distance of 300 km can be wished to be realized. In our protocol, intensity modulation is not needed, which could avoid intensity correlation caused by intensity modulators \cite{yoshino2018quantum,kang2023patterning}. With our protocol, practical side-channel-secure quantum key distribution can be wished to come true.

\textbf{Note}. During the preparation of our article, we are informed that another work of the finite-key analysis of the original SCS protocol is given \cite{jiang2023side}.
\begin{acknowledgments}
	This work has been supported by the National Key Research and Development Program of China (Grant No.2020YFA0309802), the National Natural Science Foundation of China (Grant Nos. 62171424, 62271463).
\end{acknowledgments}

\newpage
\begin{widetext}
\appendix
\section{Detailed Security Analysis\label{sec:appsec}}
In this section, we give the detailed process of our security analysis.

In the main text, we have give the state sent by Alice and Bob in the equivalent protocol, which is 
\begin{equation}
	\ket{\Phi}=\sqrt{1-P_{est}}\ket{sig}\ket{\phi}+\sqrt{P_{est}}\ket{est}\ket{\phi},\label{app:st1}
\end{equation}
\begin{equation}
		\ket{\phi}=\ket{++}_\text{a,b}\sqrt{P_{ee}}\ket{\varphi_{ee}}_{\text{A,B}}+\ket{--}_\text{a,b}\sqrt{P_{oo}}\ket{\varphi_{oo}}_{\text{A,B}}+\ket{+-}_\text{a,b}\sqrt{P_{eo}}\ket{\varphi_{eo}}_{\text{A,B}}+\ket{-+}_\text{a,b}\sqrt{P_{oe}}\ket{\varphi_{oe}}_{\text{A,B}}.\label{app:st2}
\end{equation}

The general attack of Eve and the measurement of Charlie can be treated as follows. For all pulses from Alice and Bob, Eve conducts a complete set of measurement and fabricates the clicks of a legal detection for Alice and Bob according to her measurement results. We assume that the measurement matrices of Eve are $M_e^1,M_e^2,\dots$ which satisfy $M_e^{1\dagger}M_e^1+M_e^{2\dagger}M_e^2+\dots=\mathbb{I}$. $\mathbb{I}$ is the identity matrix. Eve's measurement result can be seen as an $N$-particle state $\ket{LRO}^{N}$ sent to Alice and Bob. All particles of $\ket{LRO}^{N}$ are composed of three orthogonal states $\ket{L}$, $\ket{R}$ and $\ket{O}$. Alice and Bob measure this state to get the information of a left-click event ($\ket{L}$), a right-click event ($\ket{R}$) or a failed measurement ($\ket{O}$) for each time window. We assume that when Eve's measurement result corresponds to $M_e^x$, Eve will produce the state $\ket{LRO}^{N}_x$. Then before the measurement of Alice and Bob's ancillas, the state becomes 
\begin{equation}
	\sum_{x=1,2,\dots}\mathcal{P}\left\{M_e^x\ket{\Phi}_1\ket{\Phi}_2\dots\ket{\Phi}_N\ket{LRO}^{N}_x\right\},
\end{equation}
where $\mathcal{P}(\ket{\cdot})=\ket{\cdot}\bra{\cdot}$.

If Charlie announces a successful measurement of the left detector, we call it a ``L'' event. And if Charlie announces a successful measurement of the right detector, we call it a ``R'' event. In the following, we will give the detailed analysis for L events. And because of the symmetry of our protocol, the security of R events is identical.  

We assume that Alice and Bob measure their ancillas and the click information one by one. Note that in a real system Alice and Bob will measure all of their ancillas on the $\mathbb{Z}$ basis. However, to get the number of phase errors, we will analyze the case that Alice and Bob measure their ancillas on the $\mathbb{X}$ basis for signal windows and still on $\mathbb{Z}$ basis for parameter estimation windows. Before measuring the ancillas of the $u$-th time windows, the state becomes 
\begin{equation}
	\sum_{x=1,2,\dots} \mathcal{P}\left\{M_e^x\bigotimes_{i=1}^{u-1}(M_i^{AB}\ket{\Phi}_i)\ket{\Phi}_u\ket{\Phi}_{u+1}\dots\ket{\Phi}_N\bra{LRO}^{\otimes u-1}_{AB}\ket{LRO}^{N}_x\right\}.
\end{equation}
Here $M_e^x$ is the operator of Eve's (Charlie's) measurement, which only operates on the states sent by Alice and Bob. $M_i^{AB}$ is the measurement operator of Alice and Bob, which operated on the ancillas in the $i$-th time window. $\bra{LRO}^{\otimes u-1}_{AB}$ represents that Alice and Bob get the click information from Charlie of the first $u-1$ time windows.

Then we can get the probability of finding a phase error as an L event signal window in the $u$-th time window at the condition that the measurement of the first $u-1$ time windows have finished. This probability can be get by calculating the probability of finding a state of $\ket{++}_{\text{a,b}}$ or $\ket{--}_{\text{a,b}}$ as a signal window, which is shown as

	\begin{equation}
		\begin{aligned}
			P_{ph}^{u,L}=&\frac{\sum_{x=1,2,\dots}\left\Vert M_e^x\bigotimes_{i=1}^{u-1}(M^{AB}_i\ket{\Phi}_i)\bra{++}_{\text{a,b}}\bra{sig}{}\ket{\Phi}_u\ket{\Phi}_{u+1}\dots\ket{\Phi}_N\bra{LRO}^{\otimes u-1}\bra{L}\ket{LRO}^{N}_x\right\Vert^2}{\sum_{x=1,2,\dots}\left\Vert M_e^x\bigotimes_{i=1}^{u-1}(M^{AB}_i\ket{\Phi}_i)\ket{\Phi}_u\ket{\Phi}_{u+1}\dots\ket{\Phi}_N\bra{LRO}^{\otimes u-1}\ket{LRO}^{N}_x\right\Vert^2}\\
			&+\frac{\sum_{x=1,2,\dots}\left\Vert M_e^x\bigotimes_{i=1}^{u-1}(M^{AB}_i\ket{\Phi}_i)\bra{--}_{\text{a,b}}\bra{sig}{}\ket{\Phi}_u\ket{\Phi}_{u+1}\dots\ket{\Phi}_N\bra{LRO}^{\otimes u-1}\bra{L}\ket{LRO}^{N}_x\right\Vert^2}{\sum_{x=1,2,\dots}\left\Vert M_e^x\bigotimes_{i=1}^{u-1}(M^{AB}_i\ket{\Phi}_i)\ket{\Phi}_u\ket{\Phi}_{u+1}\dots\ket{\Phi}_N\bra{LRO}^{\otimes u-1}\ket{LRO}^{N}_x\right\Vert^2}.
		\end{aligned}
	\end{equation}

Substitute that $\ket{\Phi}_u=(\sqrt{1-P_{est}}\ket{sig}+\sqrt{P_{est}}\ket{est})\ket{\phi}_u$ into this equation, this equation becomes

	\begin{equation}
		\begin{aligned}
			P_{ph}^{u,L}=
			(1-P_{est})\frac{P_{ee}\left\Vert\ket{\varphi_{ee}}\right\Vert^2_{uL}+P_{oo}\left\Vert\ket{\varphi_{oo}}\right\Vert^2_{uL}}{P_{ee}\left\Vert\ket{\varphi_{ee}}\right\Vert^2_{u}+P_{oo}\left\Vert\ket{\varphi_{oo}}\right\Vert^2_{u}+P_{eo}\left\Vert\ket{\varphi_{eo}}\right\Vert^2_{u}+P_{oe}\left\Vert\ket{\varphi_{oe}}\right\Vert^2_{u}},
		\end{aligned}\label{eq:ph}
	\end{equation}
	where $\left\Vert\ket{\varphi_{ee}}\right\Vert^2_{uL}\equiv\sum_{x=1,2,\dots}\left\Vert M_e^x\bigotimes_{i=1}^{u-1}(M^{AB}_i\ket{\Phi}_i)\ket{\varphi_{ee}}_{\text{A,B}}^u\ket{\Phi}_{u+1}\dots\ket{\Phi}_N\bra{LRO}^{\otimes u-1}\bra{L}\ket{LRO}^{ N}_x\right\Vert^2$ and $\left\Vert\ket{\varphi_{ee}}\right\Vert^2_{u}\equiv\sum_{x=1,2,\dots}\left\Vert M_e^x\bigotimes_{i=1}^{u-1}(M^{AB}_i\ket{\Phi}_i)\ket{\varphi_{ee}}_{\text{A,B}}^u\ket{\Phi}_{u+1}\dots\ket{\Phi}_N\bra{LRO}^{\otimes u-1}\ket{LRO}^{N}_x\right\Vert^2$.

Now we consider another probability that Alice and Bob find a bit error as an L event of parameter estimation window in the $u$-th time window, which is the case that Alice and Bob find their ancillas as $\ket{01}_{\text{a,b}}$ or $\ket{10}_{\text{a,b}}$. It can also be treated as finding $(\ket{++}-\ket{--})_{\text{a,b}}/\sqrt{2}$ or $(\ket{+-}-\ket{-+})_{\text{a,b}}/\sqrt{2}$. With a similar calculation, we can get this probability as 

	\begin{equation}
		P_{est,bit}^{u,L}=\frac{1}{2}P_{est}\frac{\left\Vert\sqrt{P_{ee}}\ket{\varphi_{ee}}-\sqrt{P_{oo}}\ket{\varphi_{oo}}\right\Vert^2_{uL}+\left\Vert\sqrt{P_{eo}}\ket{\varphi_{eo}}-\sqrt{P_{oe}}\ket{\varphi_{oe}}\right\Vert^2_{uL}}{P_{ee}\left\Vert\ket{\varphi_{ee}}\right\Vert^2_{u}+P_{oo}\left\Vert\ket{\varphi_{oo}}\right\Vert^2_{u}+P_{eo}\left\Vert\ket{\varphi_{eo}}\right\Vert^2_{u}+P_{oe}\left\Vert\ket{\varphi_{oe}}\right\Vert^2_{u}}.
	\end{equation}
	
	Then we can find that,
	\begin{equation}
		\sqrt{2P_{est,bit}^{u,L}(P_{ee}\left\Vert\ket{\varphi_{ee}}\right\Vert^2_{u}+P_{oo}\left\Vert\ket{\varphi_{oo}}\right\Vert^2_{u}+P_{eo}\left\Vert\ket{\varphi_{eo}}\right\Vert^2_{u}+P_{oe}\left\Vert\ket{\varphi_{oe}}\right\Vert^2_{u})/P_{est}}\ge \sqrt{P_{ee}}\sqrt{\left\Vert\ket{\varphi_{ee}}\right\Vert^2_{uL}}-\sqrt{P_{oo}}\sqrt{\left\Vert\ket{\varphi_{oo}}\right\Vert^2_{uL}}.\label{eq:ee}
	\end{equation}
Here we have used the inequality that $\sqrt{\sum_{i}\left\Vert\ket{A_i}+\ket{B_i}\right\Vert^2}\ge\sqrt{\sum_i\left\Vert \ket{A_i}\right\Vert^2}-\sqrt{\sum_i\left\Vert \ket{B_i}\right\Vert^2}$, which equals to prove that $\sum_{i}\left\Vert\ket{A_i}+\ket{B_i}\right\Vert^2\ge \sum_i\left\Vert \ket{A_i}\right\Vert^2+\sum_i\left\Vert \ket{B_i}\right\Vert^2-2\sqrt{\sum_i\left\Vert \ket{A_i}\right\Vert^2\sum_i\left\Vert \ket{B_i}\right\Vert^2}$. Using the Cauchy-Schwarz inequality, we have 
\begin{equation}
	\begin{aligned}
\sum_i\left\Vert \ket{A_i}\right\Vert^2+\sum_i\left\Vert \ket{B_i}\right\Vert^2-2\sqrt{\sum_i\left\Vert \ket{A_i}\right\Vert^2\sum_i\left\Vert \ket{B_i}\right\Vert^2}\le& \sum_i(\left\Vert \ket{A_i}\right\Vert^2+\left\Vert \ket{B_i}\right\Vert^2- 2\left\Vert\ket{A_i}\right\Vert\left\Vert\ket{B_i}\right\Vert)\\
\le&\sum_{i}(\left\Vert \ket{A_i}\right\Vert^2+\left\Vert \ket{B_i}\right\Vert^2+\bra{A_i}\ket{B_i}+\bra{B_i}\ket{A_i})\\
=&\sum_{i}\left\Vert\ket{A_i}+\ket{B_i}\right\Vert^2.
\end{aligned}
\end{equation}
Thus the inequality is proven.

Substituting Eq. (\ref{eq:ee}) into Eq. (\ref{eq:ph}), we find the relationship between the phase errors of signal windows and the bit errors of parameter estimation windows, which is shown below.

	\begin{equation}
		\begin{aligned}
			P_{ph}^{u,L}\le& 2\frac{1-P_{est}}{P_{est}}P_{est,bit}^{u,L}+(1-P_{est})\frac{2\sqrt{2P_{est,bit}^{u,L}/P_{est}}\sqrt{P_{oo}}\sqrt{\left\Vert\ket{\varphi_{oo}}\right\Vert_{uL}^2}}{\sqrt{P_{ee}\left\Vert\ket{\varphi_{ee}}\right\Vert^2_{u}+P_{oo}\left\Vert\ket{\varphi_{oo}}\right\Vert^2_{u}+P_{eo}\left\Vert\ket{\varphi_{eo}}\right\Vert^2_{u}+P_{oe}\left\Vert\ket{\varphi_{oe}}\right\Vert^2_{u}}}\\
			&+\frac{2(1-P_{est})P_{oo}\left\Vert\ket{\varphi_{oo}}\right\Vert_{uL}^2}{P_{ee}\left\Vert\ket{\varphi_{ee}}\right\Vert^2_{u}+P_{oo}\left\Vert\ket{\varphi_{oo}}\right\Vert^2_{u}+P_{eo}\left\Vert\ket{\varphi_{eo}}\right\Vert^2_{u}+P_{oe}\left\Vert\ket{\varphi_{oe}}\right\Vert^2_{u}}.
		\end{aligned}\label{eq:phu}
	\end{equation}

Then we consider the summation of Eq.(\ref{eq:phu}) for $u$, which will be used to calculate the total number of phase errors. Using Cauchy-Schwarz inequality, we can get that

	\begin{equation}
		\begin{aligned}
			\sum_{u=1}^NP_{ph}^{u,L}\le&2\frac{1-P_{est}}{P_{est}}\sum_{u=1}^NP_{est,bit}^{u,L}+\frac{2\sqrt{2}(1-P_{est})}{\sqrt{P_{est}}}\sqrt{(\sum_{u=1}^NP_{est,bit}^{u,L})(\sum_{u=1}^N\frac{P_{oo}\left\Vert\ket{\varphi_{oo}}\right\Vert_{uL}^2}{P_{ee}\left\Vert\ket{\varphi_{ee}}\right\Vert^2_{u}+P_{oo}\left\Vert\ket{\varphi_{oo}}\right\Vert^2_{u}+P_{eo}\left\Vert\ket{\varphi_{eo}}\right\Vert^2_{u}+P_{oe}\left\Vert\ket{\varphi_{oe}}\right\Vert^2_{u}})}\\
			&+2(1-P_{est})\sum_{u=1}^N\frac{P_{oo}\left\Vert\ket{\varphi_{oo}}\right\Vert_{uL}^2}{P_{ee}\left\Vert\ket{\varphi_{ee}}\right\Vert^2_{u}+P_{oo}\left\Vert\ket{\varphi_{oo}}\right\Vert^2_{u}+P_{eo}\left\Vert\ket{\varphi_{eo}}\right\Vert^2_{u}+P_{oe}\left\Vert\ket{\varphi_{oe}}\right\Vert^2_{u}}.
		\end{aligned}\label{eq:nphl}
	\end{equation}

With a same process, we can also get the probability of a phase error as an R event. Then the total phase error rate, which is the summation of phase error rates of L and R events, is 
	\begin{equation}
\begin{aligned}
\sum_{u=1}^NP_{ph}^u\le&2\frac{1-P_{est}}{P_{est}}\sum_{u=1}^N(P_{est,bit}^{u,L}+P_{est,bit}^{u,R})\\
&+\frac{2\sqrt{2}(1-P_{est})}{\sqrt{P_{est}}}\sqrt{(\sum_{u=1}^N(P_{est,bit}^{u,L}+P_{est,bit}^{u,R}))(\sum_{u=1}^N\frac{P_{oo}(\left\Vert\ket{\varphi_{oo}}\right\Vert_{uL}^2+\left\Vert\ket{\varphi_{oo}}\right\Vert_{uR}^2)}{P_{ee}\left\Vert\ket{\varphi_{ee}}\right\Vert^2_{u}+P_{oo}\left\Vert\ket{\varphi_{oo}}\right\Vert^2_{u}+P_{eo}\left\Vert\ket{\varphi_{eo}}\right\Vert^2_{u}+P_{oe}\left\Vert\ket{\varphi_{oe}}\right\Vert^2_{u}})}\\
&+2(1-P_{est})\sum_{u=1}^N\frac{P_{oo}(\left\Vert\ket{\varphi_{oo}}\right\Vert_{uL}^2+\left\Vert\ket{\varphi_{oo}}\right\Vert_{uR}^2)}{P_{ee}\left\Vert\ket{\varphi_{ee}}\right\Vert^2_{u}+P_{oo}\left\Vert\ket{\varphi_{oo}}\right\Vert^2_{u}+P_{eo}\left\Vert\ket{\varphi_{eo}}\right\Vert^2_{u}+P_{oe}\left\Vert\ket{\varphi_{oe}}\right\Vert^2_{u}}\\
=&2\frac{1-P_{est}}{P_{est}}\sum_{u=1}^NP_{est,bit}^u\\
&+\frac{2\sqrt{2}(1-P_{est})}{\sqrt{P_{est}}}\sqrt{(\sum_{u=1}^NP_{est,bit}^u)(\sum_{u=1}^N\frac{P_{oo}(\left\Vert\ket{\varphi_{oo}}\right\Vert_{uL}^2+\left\Vert\ket{\varphi_{oo}}\right\Vert_{uR}^2)}{P_{ee}\left\Vert\ket{\varphi_{ee}}\right\Vert^2_{u}+P_{oo}\left\Vert\ket{\varphi_{oo}}\right\Vert^2_{u}+P_{eo}\left\Vert\ket{\varphi_{eo}}\right\Vert^2_{u}+P_{oe}\left\Vert\ket{\varphi_{oe}}\right\Vert^2_{u}})}\\
&+2(1-P_{est})\sum_{u=1}^N\frac{P_{oo}(\left\Vert\ket{\varphi_{oo}}\right\Vert_{uL}^2+\left\Vert\ket{\varphi_{oo}}\right\Vert_{uR}^2)}{P_{ee}\left\Vert\ket{\varphi_{ee}}\right\Vert^2_{u}+P_{oo}\left\Vert\ket{\varphi_{oo}}\right\Vert^2_{u}+P_{eo}\left\Vert\ket{\varphi_{eo}}\right\Vert^2_{u}+P_{oe}\left\Vert\ket{\varphi_{oe}}\right\Vert^2_{u}}.
\end{aligned}\label{eq:nph}
\end{equation}
Here we have used the Cauchy-Schwarz inequality when summing $P_{ph}^{u,L}$ and $P_{ph}^{u,R}$. $P_{est,bit}^u=P_{est,bit}^{u,L}+P_{est,bit}^{u,R}$ is the bit error rate as a parameter estimation window.

From the state Eq.(\ref{app:st1}, \ref{app:st2}) we can find that when measuring on $\mathbb{X}$ basis, the probability of finding a successful measurement of $\ket{--}_{\text{a,b}}$ as a signal time window at the $u$-th pulse is shown as 

With the relation that $\left\Vert\ket{\varphi_{oo}}\right\Vert_{uL}^2+\left\Vert\ket{\varphi_{oo}}\right\Vert_{uR}^2+\left\Vert\ket{\varphi_{oo}}\right\Vert_{uO}^2=\left\Vert\ket{\varphi_{oo}}\right\Vert_{u}^2$, we can get the inequality that

	\begin{equation}
		\begin{aligned}
		(1-P_{est})\frac{P_{oo}(\left\Vert\ket{\varphi_{oo}}\right\Vert_{uL}^2+\left\Vert\ket{\varphi_{oo}}\right\Vert_{uR}^2)}{P_{ee}\left\Vert\ket{\varphi_{ee}}\right\Vert^2_{u}+P_{oo}\left\Vert\ket{\varphi_{oo}}\right\Vert^2_{u}+P_{eo}\left\Vert\ket{\varphi_{eo}}\right\Vert^2_{u}+P_{oe}\left\Vert\ket{\varphi_{oe}}\right\Vert^2_{u}}\\
		\le (1-P_{est})\frac{P_{oo}\left\Vert\ket{\varphi_{oo}}\right\Vert_{u}^2}{P_{ee}\left\Vert\ket{\varphi_{ee}}\right\Vert^2_{u}+P_{oo}\left\Vert\ket{\varphi_{oo}}\right\Vert^2_{u}+P_{eo}\left\Vert\ket{\varphi_{eo}}\right\Vert^2_{u}+P_{oe}\left\Vert\ket{\varphi_{oe}}\right\Vert^2_{u}},
		\end{aligned}
	\end{equation}
which is the probability of finding a signal state $\ket{--}_{\text{a,b}}$ in the $u$-th time window.

With Kato's concentration inequality (detailed in Appendix \ref{bound}), we can relate this value to the number of measuring result.

	\begin{equation}
		\begin{aligned}
			(1-P_{est})\sum_{u=1}^N\frac{P_{oo}\left\Vert\ket{\varphi_{oo}}\right\Vert_{u}^2}{P_{ee}\left\Vert\ket{\varphi_{ee}}\right\Vert^2_{u}+P_{oo}\left\Vert\ket{\varphi_{oo}}\right\Vert^2_{u}+P_{eo}\left\Vert\ket{\varphi_{eo}}\right\Vert^2_{u}+P_{oe}\left\Vert\ket{\varphi_{oe}}\right\Vert^2_{u}}\le  \text{U}_e^{\epsilon^2}(N_{sig}^{--}).
		\end{aligned}
	\end{equation}

where $N_{sig}^{--}$ is the number of events that Alice and Bob find $\ket{--}_{\text{a,b}}$ in signal windows no matter Charlie declares a successful measurement or not, which is also the number of state $\ket{\varphi_{oo}}_{\text{A,B}}$ produced by Alice and Bob. 

Note that the order of measurements do not influence the measurement result, which means the two kinds of measurements below are equivalent.
\begin{enumerate}
	\item 
	Alice and Bob measure their ancillas and the click information of the first time window. Then they do the same measurement to the next time window. They measure all times windows one by one.
	\item 
	Alice and Bob measure all ancillas of all time windows. Then they measure all click information from Charlie (Eve).
\end{enumerate}
Results of these two kinds of measurements obey a same probability distribution. Thus these two kinds of measurements are equivalent, which means the analyses from the two cases hold at the same time. If Alice and Bob use the second kind of measurement, we can easily get that the random variables of finding a state $\ket{sig}\ket{--}_\text{a,b}$ of every time window are independent. Thus we can use the Chernoff bound of independent random variables to bound $N_{sig}^{--}$, which is 
\begin{equation}
	N_{sig}^{--}\le\text{C}_U^{\epsilon^2}(N(1-P_{est})\bar{P}_{oo}).
\end{equation}
The detail of Chernoff bound can be seen in Appendix \ref{bound}. Here $\bar{P}_{oo}$ is the upper bound of the probability for producing a $\ket{--}_{\text{a,b}}$ for every time window, which can be calculated as follows.

	\begin{equation}
		\begin{aligned}
			\sqrt{P_{oo}}\ket{\varphi_{oo}}=\frac{1}{4}(\ket{\alpha',\alpha'}+\ket{-\alpha',-\alpha'}-\ket{\alpha',-\alpha'}-\ket{-\alpha',\alpha'}),
		\end{aligned}
	\end{equation}
	\begin{equation}
		P_{oo}=\frac{1}{16}(\bra{\alpha',\alpha'}+\bra{-\alpha',-\alpha'}-\bra{\alpha',-\alpha'}-\bra{-\alpha',\alpha'})(\ket{\alpha',\alpha'}+\ket{-\alpha',-\alpha'}-\ket{\alpha',-\alpha'}-\ket{-\alpha',\alpha'}).
	\end{equation}
	Note that the state of Alice and Bob can be different and the intensity of $\ket{\alpha'}$ and $\ket{-\alpha'}$ can be different. We have the definition below.
	\begin{equation}
		\begin{aligned}
			\ket{\alpha'}_{A}=\sqrt{P_{0A}^{+}}\ket{0}+\sqrt{1-P_{0A}^{+}}\ket{\varphi}_A,\\
			\ket{-\alpha'}_{A}=\sqrt{P_{0A}^{-}}\ket{0}+\sqrt{1-P_{0A}^{-}}\ket{\psi}_A,\\
			\ket{\alpha'}_{B}=\sqrt{P_{0B}^{+}}\ket{0}+\sqrt{1-P_{0B}^{+}}\ket{\varphi}_B,\\
			\ket{-\alpha'}_{B}=\sqrt{P_{0B}^{-}}\ket{0}+\sqrt{1-P_{0B}^{-}}\ket{\psi}_B,\\
		\end{aligned}
	\end{equation}
	\begin{equation}
		\braket{\varphi}{\psi}_A=X_A;\braket{\varphi}{\psi}_B=X_B.
	\end{equation}
	Thus we can find that 
	\begin{equation}
		\begin{aligned}
			P_{oo}&=\frac{1}{16}\left(\sqrt{(1-P_{0A}^+)(1-P_{0A}^-)}(X_A+X_A^*)-2\left(1-\sqrt{P_{0A}^+P_{0A}^-}\right)\right)\left(\sqrt{(1-P_{0B}^+)(1-P_{0B}^-)}(X_B+X_B^*)-2\left(1-\sqrt{P_{0B}^+P_{0B}^-}\right)\right)\\
			&\le (1-\underline{P}_{0})^2.
		\end{aligned}
	\end{equation}
	The inequality is obvious because the maximum is gotten on the condition that $X_A=X_B=-1$, then this equation is decreasing for $P_{0A}^+$, $P_{0A}^-$, $P_{0B}^+$, and $P_{0B}^-$. $\underline{P}_{0}$ is the lower bound of $P_{0}$, which is $\text{e}^{-\mu}$ for a coherent state ($\mu$ is the maximum average photon number of every pulse).

Now the rest unknown items of Eq.(\ref{eq:nph}) is $\sum_{u=1}^NP_{est,bit}^u$, which can be easily related to the number of bit errors in estimation windows using Kato's inequality.

\begin{equation}
	\sum_{u=1}^NP_{est,bit}^u\le \text{U}_e^{\epsilon^2}(N_{est,bit})\label{eq:Nbit}.
\end{equation}
Details of Kato's inequality can be seen in Appendix \ref{bound}. Here $N_{est,bit}$ is the number of bit errors of estimation windows for L or R events when Alice and Bob detect on $\mathbb{Z}$ basis, which can be counted in realistic experiments. 

Thus from Eq.(\ref{eq:nph}) to (\ref{eq:Nbit}), we can get the final result of the phase errors with another Kato's inequality.
\begin{equation}
		\begin{aligned}
			\sum_{u=1}^NP_{ph}^u\le2\frac{1-P_{est}}{P_{est}}\text{U}^{\epsilon^2}_{e}(N_{est,bit})
			+\frac{2\sqrt{2}\sqrt{(1-P_{est})}}{\sqrt{P_{est}}}\sqrt{\text{U}^{\epsilon^2}_{e}(N_{est,bit})\text{U}^{\epsilon^2}_{e}\left[\text{C}_{U}^{\epsilon^2}\left(N(1-P_{est})(1-\text{e}^{-\mu})^2\right)\right]}
			\\+2\text{U}^{\epsilon^2}_{e}\left[\text{C}_{U}^{\epsilon^2}(N(1-P_{est})(1-\text{e}^{-\mu})^2)\right],
		\end{aligned}\label{eq:sumph}
\end{equation}
	\begin{equation}
		N_{ph}\le \text{U}_{m}^{\epsilon^2}(\sum_{u=1}^NP_{ph}^u)\equiv \bar{N}_{ph}.\label{eq:uph}
	\end{equation}
	
	Realizing that we have used three Kato's inequalities and one Chernoff bound, we have set the failure parameters of these four inequalities to be the same value $\epsilon^2$. The failure parameter of these four inequalities is $4\epsilon^2$. 
	
	Considering the secure parameter is $\epsilon_{sec}=2\epsilon'+\frac{1}{2}\sqrt{2^{l-H_{min}^{\epsilon'}(A|E')}}=2\epsilon'+\tilde{\epsilon}$ for the length of key bits $l=H_{min}^{\epsilon'}(A|E')-2\log_2\frac{1}{2\tilde{\epsilon}}$. Here $\epsilon'=\sqrt{4\epsilon^2}=2\epsilon$. Then the final key length becomes \cite{renner2008security}
\begin{equation}
	l=N_{sig}(1-H_2(\frac{\bar{N}_{ph}}{N_{sig}})-fH_2(e_{bit}))-2\log_2\frac{1}{2\tilde\epsilon}-\log_2\frac{2}{\epsilon_{cor}}.
\end{equation}
The total secure parameter is $\epsilon_{tot}=4\epsilon+\tilde\epsilon+\epsilon_{cor}$.

\section{Kato's inequality and Chernoff bound\label{bound}}
In this section, we will simply introduce Kato's inequality and Chernoff bound used in our analysis.
\subsection{Kato's inequality}
Kato's inequality \cite{kato2020concentration} is an improved version of Azuma's inequality \cite{azuma1967weighted}, which has been widely used in the security analysis of quantum key distribution.
\\

\textbf{Kato's inequality.} Let $\{X_m\}$ be a list of random variables, and $\mathcal{F}_m$ be the measurement result of the random variables $X_1,X_2,\dots,X_m$. Suppose that $0\le X_m\le1$ for all m. In this case, for any $n\in \mathbb{N}$, $a\in\mathbb{R}$ and $b\in\mathbb{R}_{\ge0}$,
\begin{equation}
	P\left(\sum_{m=1}^n(E(X_m|\mathcal{F}_{m-1})-X_m)\ge(b+a(2\frac{\sum_{m=1}^n X_m}{n}-1))\sqrt{n}\right)\le\exp(-\frac{2(b^2-a^2)}{(1+\frac{4a}{3\sqrt{n}})^2})
\end{equation}
holds.
\\

We denote that $\Lambda=\sum_m^nX_m$, then we can get that
\begin{equation}
	\text{Pr}\left(\sum_{m=1}^nE(X_m|\mathcal{F}_{m-1})-\Lambda\ge\left(b+a(\frac{2\Lambda}{n}-1)\right)\sqrt{n}\right)\le\exp\left(-\frac{2(b^2-a^2)}{(1+\frac{4a}{3\sqrt{n}})^2}\right)\label{concen:1},
\end{equation}
and
\begin{equation}
	\text{Pr}\left(\Lambda-\sum_{m=1}^nE(X_m|\mathcal{F}_{m-1})\ge\left(b+a(\frac{2\Lambda}{n}-1)\right)\sqrt{n}\right)\le\exp\left(-\frac{2(b^2-a^2)}{(1-\frac{4a}{3\sqrt{n}})^2}\right)\label{concen:2},
\end{equation}
by replacing $X_m\to 1-X_m$ and $a\to -a$. \cite{curras2021tight}

For Eq. (\ref{concen:1}), to get a tight bound, we let $\exp(-\frac{2(b^2-a^2)}{(1+\frac{4a}{3\sqrt{n}})^2})=\epsilon$ and solve $\min[\left(b+a(\frac{2\Lambda}{n}-1)\right)]$. The optimal value of $a$ and $b$ are
\begin{equation}
	a_1=\frac{3\left(72\sqrt{n}\Lambda(n-\Lambda)\ln\epsilon-16n^{3/2}\ln^2\epsilon+9\sqrt{2}(n-2\Lambda)\sqrt{-n^2\ln\epsilon(9\Lambda(n-\Lambda)-2n\ln\epsilon)}\right)}{4(9n-8\ln\epsilon)(9\Lambda(n-\Lambda)-2n\ln\epsilon)},
\end{equation}
\begin{equation}
	b_1=\frac{\sqrt{18a^2n-(16a^2+24a\sqrt{n}+9n)\ln\epsilon}}{3\sqrt{2n}}.
\end{equation}

Then for known measurement result of random variables (known $\Lambda$), the upper bound of mathematical expectations can be get by 
\begin{equation}
	\sum_{m=1}^nE(X_m|\mathcal{F}_{m-1})\le\Lambda+\left(b_1+a_1(\frac{2\Lambda}{n}-1)\right)\sqrt{n}\equiv \text{U}_e^\epsilon(\Lambda).
\end{equation}
And with known expectations, the lower bound of $\Lambda$ is 
\begin{equation}
	\Lambda\ge\frac{\sum_{m=1}^nE(X_m|\mathcal{F}_{m-1})-(b_1-a_1)\sqrt{n}}{1+\frac{2a_1}{\sqrt{n}}}\equiv \text{L}_m^\epsilon(\sum_{m=1}^nE(X_m|\mathcal{F}_{m-1})).
\end{equation}

For Eq. (\ref{concen:2}), with a similar process, we can get 
\begin{equation}
	a_2=\frac{-3\left(72\sqrt{n}\Lambda(n-\Lambda)\ln\epsilon-16n^{3/2}\ln^2\epsilon-9\sqrt{2}(n-2\Lambda)\sqrt{-n^2\ln\epsilon(9\Lambda(n-\Lambda)-2n\ln\epsilon)}\right)}{4(9n-8\ln\epsilon)(9\Lambda(n-\Lambda)-2n\ln\epsilon)},
\end{equation}
\begin{equation}
	b_2=\frac{\sqrt{18a^2n-(16a^2-24a\sqrt{n}+9n)\ln\epsilon}}{3\sqrt{2n}},
\end{equation}

\begin{equation}
	\sum_{m=1}^nE(X_m|\mathcal{F}_{m-1})\ge\Lambda-\left(b_2+a_2(\frac{2\Lambda}{n}-1)\right)\sqrt{n}\equiv \text{L}_e^\epsilon(\Lambda),
\end{equation}

\begin{equation}
	\Lambda\le\frac{\sum_{m=1}^nE(X_m|\mathcal{F}_{m-1})+(b_2-a_2)\sqrt{n}}{1-\frac{2a_2}{\sqrt{n}}}\equiv \text{U}_m^\epsilon(\sum_{m=1}^nE(X_m|\mathcal{F}_{m-1})).
\end{equation}

\subsection{Chernoff bound}
In our security analysis, we use the Chernoff bound \cite{mitzenmacher2017probability} to get the upper bound of measurement result for independent random variables.
\\

\textbf{Multiplicative Chernoff bound.} Suppose $X_1,X_2,\dots,X_n$ are independent random variables taking values in $\{0,1\}$. Let $X=X_1+X_2+\dots+X_n$ and $\mu=E[X]$ is its expectation. Then for $\delta>0$,
\begin{equation}
	P(X>(1+\delta)\mu)\le(\frac{\text{e}^{\delta}}{(1+\delta)^{1+\delta}})^\mu,
\end{equation}
and
\begin{equation}
	P(X<(1-\delta)\mu)\le(\frac{\text{e}^{-\delta}}{(1-\delta)^{1-\delta}})^\mu
\end{equation}
\\

With the equality that $\frac{2\delta}{2+\delta}\le\ln(1+\delta)$, we can get the inequality we used.
\begin{equation}
	P(X\ge(1+\delta)\mu)\le\text{e}^{-\delta^2\mu/(2+\delta)},\ \ \delta\ge0.
\end{equation}
We let $\text{e}^{-\delta^2\mu/(2+\delta)}=\epsilon$, then
\begin{equation}
	\text{C}_U^\epsilon(\mu)\equiv (1+\delta)\mu
\end{equation}
\begin{equation}
	\delta=\frac{\ln\frac{1}{\epsilon}+\sqrt{(\ln\frac{1}{\epsilon})^2+8\mu\ln\frac{1}{\epsilon}}}{2\mu}
\end{equation}

\end{widetext}

\bibliography{refe.bib}

\begin{thebibliography}{49}%
\makeatletter
\providecommand \@ifxundefined [1]{%
 \@ifx{#1\undefined}
}%
\providecommand \@ifnum [1]{%
 \ifnum #1\expandafter \@firstoftwo
 \else \expandafter \@secondoftwo
 \fi
}%
\providecommand \@ifx [1]{%
 \ifx #1\expandafter \@firstoftwo
 \else \expandafter \@secondoftwo
 \fi
}%
\providecommand \natexlab [1]{#1}%
\providecommand \enquote  [1]{``#1''}%
\providecommand \bibnamefont  [1]{#1}%
\providecommand \bibfnamefont [1]{#1}%
\providecommand \citenamefont [1]{#1}%
\providecommand \href@noop [0]{\@secondoftwo}%
\providecommand \href [0]{\begingroup \@sanitize@url \@href}%
\providecommand \@href[1]{\@@startlink{#1}\@@href}%
\providecommand \@@href[1]{\endgroup#1\@@endlink}%
\providecommand \@sanitize@url [0]{\catcode `\\12\catcode `\$12\catcode
  `\&12\catcode `\#12\catcode `\^12\catcode `\_12\catcode `\%12\relax}%
\providecommand \@@startlink[1]{}%
\providecommand \@@endlink[0]{}%
\providecommand \url  [0]{\begingroup\@sanitize@url \@url }%
\providecommand \@url [1]{\endgroup\@href {#1}{\urlprefix }}%
\providecommand \urlprefix  [0]{URL }%
\providecommand \Eprint [0]{\href }%
\providecommand \doibase [0]{https://doi.org/}%
\providecommand \selectlanguage [0]{\@gobble}%
\providecommand \bibinfo  [0]{\@secondoftwo}%
\providecommand \bibfield  [0]{\@secondoftwo}%
\providecommand \translation [1]{[#1]}%
\providecommand \BibitemOpen [0]{}%
\providecommand \bibitemStop [0]{}%
\providecommand \bibitemNoStop [0]{.\EOS\space}%
\providecommand \EOS [0]{\spacefactor3000\relax}%
\providecommand \BibitemShut  [1]{\csname bibitem#1\endcsname}%
\let\auto@bib@innerbib\@empty
\bibitem [{\citenamefont {Bennett}\ and\ \citenamefont
  {Brassard}(1984)}]{bennett1984quantum}%
  \BibitemOpen
  \bibfield  {author} {\bibinfo {author} {\bibfnamefont {C.~H.}\ \bibnamefont
  {Bennett}}\ and\ \bibinfo {author} {\bibfnamefont {G.}~\bibnamefont
  {Brassard}},\ }\bibfield  {title} {\bibinfo {title} {Quantum cryptography:
  public key distribution and coin tossing int},\ }in\ \href@noop {} {\emph
  {\bibinfo {booktitle} {Conf. on Computers, Systems and Signal Processing
  (Bangalore, India, Dec. 1984)}}}\ (\bibinfo {year} {1984})\ pp.\ \bibinfo
  {pages} {175--179}\BibitemShut {NoStop}%
\bibitem [{\citenamefont {Renner}(2008)}]{renner2008security}%
  \BibitemOpen
  \bibfield  {author} {\bibinfo {author} {\bibfnamefont {R.}~\bibnamefont
  {Renner}},\ }\bibfield  {title} {\bibinfo {title} {Security of quantum key
  distribution},\ }\href@noop {} {\bibfield  {journal} {\bibinfo  {journal}
  {International Journal of Quantum Information}\ }\textbf {\bibinfo {volume}
  {6}},\ \bibinfo {pages} {1} (\bibinfo {year} {2008})}\BibitemShut {NoStop}%
\bibitem [{\citenamefont {Brassard}\ \emph {et~al.}(2000)\citenamefont
  {Brassard}, \citenamefont {L{\"u}tkenhaus}, \citenamefont {Mor},\ and\
  \citenamefont {Sanders}}]{brassard2000limitations}%
  \BibitemOpen
  \bibfield  {author} {\bibinfo {author} {\bibfnamefont {G.}~\bibnamefont
  {Brassard}}, \bibinfo {author} {\bibfnamefont {N.}~\bibnamefont
  {L{\"u}tkenhaus}}, \bibinfo {author} {\bibfnamefont {T.}~\bibnamefont
  {Mor}},\ and\ \bibinfo {author} {\bibfnamefont {B.~C.}\ \bibnamefont
  {Sanders}},\ }\bibfield  {title} {\bibinfo {title} {Limitations on practical
  quantum cryptography},\ }\href@noop {} {\bibfield  {journal} {\bibinfo
  {journal} {Physical review letters}\ }\textbf {\bibinfo {volume} {85}},\
  \bibinfo {pages} {1330} (\bibinfo {year} {2000})}\BibitemShut {NoStop}%
\bibitem [{\citenamefont {L{\"u}tkenhaus}\ and\ \citenamefont
  {Jahma}(2002)}]{lutkenhaus2002quantum}%
  \BibitemOpen
  \bibfield  {author} {\bibinfo {author} {\bibfnamefont {N.}~\bibnamefont
  {L{\"u}tkenhaus}}\ and\ \bibinfo {author} {\bibfnamefont {M.}~\bibnamefont
  {Jahma}},\ }\bibfield  {title} {\bibinfo {title} {Quantum key distribution
  with realistic states: photon-number statistics in the photon-number
  splitting attack},\ }\href@noop {} {\bibfield  {journal} {\bibinfo  {journal}
  {New Journal of Physics}\ }\textbf {\bibinfo {volume} {4}},\ \bibinfo {pages}
  {44} (\bibinfo {year} {2002})}\BibitemShut {NoStop}%
\bibitem [{\citenamefont {Qi}\ \emph {et~al.}(2007)\citenamefont {Qi},
  \citenamefont {Fung}, \citenamefont {Lo},\ and\ \citenamefont
  {Ma}}]{qi2007time}%
  \BibitemOpen
  \bibfield  {author} {\bibinfo {author} {\bibfnamefont {B.}~\bibnamefont
  {Qi}}, \bibinfo {author} {\bibfnamefont {C.~H.~F.}\ \bibnamefont {Fung}},
  \bibinfo {author} {\bibfnamefont {H.~K.}\ \bibnamefont {Lo}},\ and\ \bibinfo
  {author} {\bibfnamefont {X.}~\bibnamefont {Ma}},\ }\bibfield  {title}
  {\bibinfo {title} {Time-shift attack in practical quantum cryptosystems},\
  }\href@noop {} {\bibfield  {journal} {\bibinfo  {journal} {Quantum
  Information and Computation}\ } (\bibinfo {year} {2007})}\BibitemShut
  {NoStop}%
\bibitem [{\citenamefont {Inamori}\ \emph {et~al.}(2007)\citenamefont
  {Inamori}, \citenamefont {L{\"u}tkenhaus},\ and\ \citenamefont
  {Mayers}}]{inamori2007unconditional}%
  \BibitemOpen
  \bibfield  {author} {\bibinfo {author} {\bibfnamefont {H.}~\bibnamefont
  {Inamori}}, \bibinfo {author} {\bibfnamefont {N.}~\bibnamefont
  {L{\"u}tkenhaus}},\ and\ \bibinfo {author} {\bibfnamefont {D.}~\bibnamefont
  {Mayers}},\ }\bibfield  {title} {\bibinfo {title} {Unconditional security of
  practical quantum key distribution},\ }\href@noop {} {\bibfield  {journal}
  {\bibinfo  {journal} {The European Physical Journal D}\ }\textbf {\bibinfo
  {volume} {41}},\ \bibinfo {pages} {599} (\bibinfo {year} {2007})}\BibitemShut
  {NoStop}%
\bibitem [{\citenamefont {Wang}\ \emph {et~al.}(2008)\citenamefont {Wang},
  \citenamefont {Peng}, \citenamefont {Zhang}, \citenamefont {Yang},\ and\
  \citenamefont {Pan}}]{wang2008general}%
  \BibitemOpen
  \bibfield  {author} {\bibinfo {author} {\bibfnamefont {X.-B.}\ \bibnamefont
  {Wang}}, \bibinfo {author} {\bibfnamefont {C.-Z.}\ \bibnamefont {Peng}},
  \bibinfo {author} {\bibfnamefont {J.}~\bibnamefont {Zhang}}, \bibinfo
  {author} {\bibfnamefont {L.}~\bibnamefont {Yang}},\ and\ \bibinfo {author}
  {\bibfnamefont {J.-W.}\ \bibnamefont {Pan}},\ }\bibfield  {title} {\bibinfo
  {title} {General theory of decoy-state quantum cryptography with source
  errors},\ }\href@noop {} {\bibfield  {journal} {\bibinfo  {journal} {Physical
  Review A}\ }\textbf {\bibinfo {volume} {77}},\ \bibinfo {pages} {042311}
  (\bibinfo {year} {2008})}\BibitemShut {NoStop}%
\bibitem [{\citenamefont {Zhao}\ \emph {et~al.}(2008)\citenamefont {Zhao},
  \citenamefont {Fung}, \citenamefont {Qi}, \citenamefont {Chen},\ and\
  \citenamefont {Lo}}]{PhysRevA.78.042333}%
  \BibitemOpen
  \bibfield  {author} {\bibinfo {author} {\bibfnamefont {Y.}~\bibnamefont
  {Zhao}}, \bibinfo {author} {\bibfnamefont {C.-H.~F.}\ \bibnamefont {Fung}},
  \bibinfo {author} {\bibfnamefont {B.}~\bibnamefont {Qi}}, \bibinfo {author}
  {\bibfnamefont {C.}~\bibnamefont {Chen}},\ and\ \bibinfo {author}
  {\bibfnamefont {H.-K.}\ \bibnamefont {Lo}},\ }\bibfield  {title} {\bibinfo
  {title} {Quantum hacking: Experimental demonstration of time-shift attack
  against practical quantum-key-distribution systems},\ }\href
  {https://doi.org/10.1103/PhysRevA.78.042333} {\bibfield  {journal} {\bibinfo
  {journal} {Phys. Rev. A}\ }\textbf {\bibinfo {volume} {78}},\ \bibinfo
  {pages} {042333} (\bibinfo {year} {2008})}\BibitemShut {NoStop}%
\bibitem [{\citenamefont {Scarani}\ \emph {et~al.}(2009)\citenamefont
  {Scarani}, \citenamefont {Bechmann-Pasquinucci}, \citenamefont {Cerf},
  \citenamefont {Du{\v{s}}ek}, \citenamefont {L{\"u}tkenhaus},\ and\
  \citenamefont {Peev}}]{scarani2009security}%
  \BibitemOpen
  \bibfield  {author} {\bibinfo {author} {\bibfnamefont {V.}~\bibnamefont
  {Scarani}}, \bibinfo {author} {\bibfnamefont {H.}~\bibnamefont
  {Bechmann-Pasquinucci}}, \bibinfo {author} {\bibfnamefont {N.~J.}\
  \bibnamefont {Cerf}}, \bibinfo {author} {\bibfnamefont {M.}~\bibnamefont
  {Du{\v{s}}ek}}, \bibinfo {author} {\bibfnamefont {N.}~\bibnamefont
  {L{\"u}tkenhaus}},\ and\ \bibinfo {author} {\bibfnamefont {M.}~\bibnamefont
  {Peev}},\ }\bibfield  {title} {\bibinfo {title} {The security of practical
  quantum key distribution},\ }\href@noop {} {\bibfield  {journal} {\bibinfo
  {journal} {Reviews of modern physics}\ }\textbf {\bibinfo {volume} {81}},\
  \bibinfo {pages} {1301} (\bibinfo {year} {2009})}\BibitemShut {NoStop}%
\bibitem [{\citenamefont {Makarov}(2009)}]{makarov2009controlling}%
  \BibitemOpen
  \bibfield  {author} {\bibinfo {author} {\bibfnamefont {V.}~\bibnamefont
  {Makarov}},\ }\bibfield  {title} {\bibinfo {title} {Controlling passively
  quenched single photon detectors by bright light},\ }\href@noop {} {\bibfield
   {journal} {\bibinfo  {journal} {New Journal of Physics}\ }\textbf {\bibinfo
  {volume} {11}},\ \bibinfo {pages} {065003} (\bibinfo {year}
  {2009})}\BibitemShut {NoStop}%
\bibitem [{\citenamefont {Lydersen}\ \emph {et~al.}(2010)\citenamefont
  {Lydersen}, \citenamefont {Wiechers}, \citenamefont {Wittmann}, \citenamefont
  {Elser}, \citenamefont {Skaar},\ and\ \citenamefont
  {Makarov}}]{lydersen2010hacking}%
  \BibitemOpen
  \bibfield  {author} {\bibinfo {author} {\bibfnamefont {L.}~\bibnamefont
  {Lydersen}}, \bibinfo {author} {\bibfnamefont {C.}~\bibnamefont {Wiechers}},
  \bibinfo {author} {\bibfnamefont {C.}~\bibnamefont {Wittmann}}, \bibinfo
  {author} {\bibfnamefont {D.}~\bibnamefont {Elser}}, \bibinfo {author}
  {\bibfnamefont {J.}~\bibnamefont {Skaar}},\ and\ \bibinfo {author}
  {\bibfnamefont {V.}~\bibnamefont {Makarov}},\ }\bibfield  {title} {\bibinfo
  {title} {Hacking commercial quantum cryptography systems by tailored bright
  illumination},\ }\href@noop {} {\bibfield  {journal} {\bibinfo  {journal}
  {Nature photonics}\ }\textbf {\bibinfo {volume} {4}},\ \bibinfo {pages} {686}
  (\bibinfo {year} {2010})}\BibitemShut {NoStop}%
\bibitem [{\citenamefont {Wiechers}\ \emph {et~al.}(2011)\citenamefont
  {Wiechers}, \citenamefont {Lydersen}, \citenamefont {Wittmann}, \citenamefont
  {Elser}, \citenamefont {Skaar}, \citenamefont {Marquardt}, \citenamefont
  {Makarov},\ and\ \citenamefont {Leuchs}}]{wiechers2011after}%
  \BibitemOpen
  \bibfield  {author} {\bibinfo {author} {\bibfnamefont {C.}~\bibnamefont
  {Wiechers}}, \bibinfo {author} {\bibfnamefont {L.}~\bibnamefont {Lydersen}},
  \bibinfo {author} {\bibfnamefont {C.}~\bibnamefont {Wittmann}}, \bibinfo
  {author} {\bibfnamefont {D.}~\bibnamefont {Elser}}, \bibinfo {author}
  {\bibfnamefont {J.}~\bibnamefont {Skaar}}, \bibinfo {author} {\bibfnamefont
  {C.}~\bibnamefont {Marquardt}}, \bibinfo {author} {\bibfnamefont
  {V.}~\bibnamefont {Makarov}},\ and\ \bibinfo {author} {\bibfnamefont
  {G.}~\bibnamefont {Leuchs}},\ }\bibfield  {title} {\bibinfo {title}
  {After-gate attack on a quantum cryptosystem},\ }\href@noop {} {\bibfield
  {journal} {\bibinfo  {journal} {New Journal of Physics}\ }\textbf {\bibinfo
  {volume} {13}},\ \bibinfo {pages} {013043} (\bibinfo {year}
  {2011})}\BibitemShut {NoStop}%
\bibitem [{\citenamefont {Lu}\ \emph {et~al.}(2023)\citenamefont {Lu},
  \citenamefont {Ye}, \citenamefont {Wang}, \citenamefont {Wang}, \citenamefont
  {Yin}, \citenamefont {Wang}, \citenamefont {Huang}, \citenamefont {Chen},
  \citenamefont {He}, \citenamefont {Fan-Yuan} \emph {et~al.}}]{lu2023hacking}%
  \BibitemOpen
  \bibfield  {author} {\bibinfo {author} {\bibfnamefont {F.-Y.}\ \bibnamefont
  {Lu}}, \bibinfo {author} {\bibfnamefont {P.}~\bibnamefont {Ye}}, \bibinfo
  {author} {\bibfnamefont {Z.-H.}\ \bibnamefont {Wang}}, \bibinfo {author}
  {\bibfnamefont {S.}~\bibnamefont {Wang}}, \bibinfo {author} {\bibfnamefont
  {Z.-Q.}\ \bibnamefont {Yin}}, \bibinfo {author} {\bibfnamefont
  {R.}~\bibnamefont {Wang}}, \bibinfo {author} {\bibfnamefont {X.-J.}\
  \bibnamefont {Huang}}, \bibinfo {author} {\bibfnamefont {W.}~\bibnamefont
  {Chen}}, \bibinfo {author} {\bibfnamefont {D.-Y.}\ \bibnamefont {He}},
  \bibinfo {author} {\bibfnamefont {G.-J.}\ \bibnamefont {Fan-Yuan}}, \emph
  {et~al.},\ }\bibfield  {title} {\bibinfo {title} {Hacking
  measurement-device-independent quantum key distribution},\ }\href@noop {}
  {\bibfield  {journal} {\bibinfo  {journal} {Optica}\ }\textbf {\bibinfo
  {volume} {10}},\ \bibinfo {pages} {520} (\bibinfo {year} {2023})}\BibitemShut
  {NoStop}%
\bibitem [{\citenamefont {Mayers}\ and\ \citenamefont
  {Yao}(1998)}]{mayers1998quantum}%
  \BibitemOpen
  \bibfield  {author} {\bibinfo {author} {\bibfnamefont {D.}~\bibnamefont
  {Mayers}}\ and\ \bibinfo {author} {\bibfnamefont {A.}~\bibnamefont {Yao}},\
  }\bibfield  {title} {\bibinfo {title} {Quantum cryptography with imperfect
  apparatus},\ }in\ \href@noop {} {\emph {\bibinfo {booktitle} {Proceedings
  39th Annual Symposium on Foundations of Computer Science (Cat. No.
  98CB36280)}}}\ (\bibinfo {organization} {IEEE},\ \bibinfo {year} {1998})\
  pp.\ \bibinfo {pages} {503--509}\BibitemShut {NoStop}%
\bibitem [{\citenamefont {Ac{\'\i}n}\ \emph {et~al.}(2007)\citenamefont
  {Ac{\'\i}n}, \citenamefont {Brunner}, \citenamefont {Gisin}, \citenamefont
  {Massar}, \citenamefont {Pironio},\ and\ \citenamefont
  {Scarani}}]{acin2007device}%
  \BibitemOpen
  \bibfield  {author} {\bibinfo {author} {\bibfnamefont {A.}~\bibnamefont
  {Ac{\'\i}n}}, \bibinfo {author} {\bibfnamefont {N.}~\bibnamefont {Brunner}},
  \bibinfo {author} {\bibfnamefont {N.}~\bibnamefont {Gisin}}, \bibinfo
  {author} {\bibfnamefont {S.}~\bibnamefont {Massar}}, \bibinfo {author}
  {\bibfnamefont {S.}~\bibnamefont {Pironio}},\ and\ \bibinfo {author}
  {\bibfnamefont {V.}~\bibnamefont {Scarani}},\ }\bibfield  {title} {\bibinfo
  {title} {Device-independent security of quantum cryptography against
  collective attacks},\ }\href@noop {} {\bibfield  {journal} {\bibinfo
  {journal} {Physical Review Letters}\ }\textbf {\bibinfo {volume} {98}},\
  \bibinfo {pages} {230501} (\bibinfo {year} {2007})}\BibitemShut {NoStop}%
\bibitem [{\citenamefont {Lo}\ \emph {et~al.}(2012)\citenamefont {Lo},
  \citenamefont {Curty},\ and\ \citenamefont {Qi}}]{lo2012measurement}%
  \BibitemOpen
  \bibfield  {author} {\bibinfo {author} {\bibfnamefont {H.-K.}\ \bibnamefont
  {Lo}}, \bibinfo {author} {\bibfnamefont {M.}~\bibnamefont {Curty}},\ and\
  \bibinfo {author} {\bibfnamefont {B.}~\bibnamefont {Qi}},\ }\bibfield
  {title} {\bibinfo {title} {Measurement-device-independent quantum key
  distribution},\ }\href@noop {} {\bibfield  {journal} {\bibinfo  {journal}
  {Phys. Rev. Lett.}\ }\textbf {\bibinfo {volume} {108}},\ \bibinfo {pages}
  {130503} (\bibinfo {year} {2012})}\BibitemShut {NoStop}%
\bibitem [{\citenamefont {Braunstein}\ and\ \citenamefont
  {Pirandola}(2012)}]{braunstein2012side}%
  \BibitemOpen
  \bibfield  {author} {\bibinfo {author} {\bibfnamefont {S.~L.}\ \bibnamefont
  {Braunstein}}\ and\ \bibinfo {author} {\bibfnamefont {S.}~\bibnamefont
  {Pirandola}},\ }\bibfield  {title} {\bibinfo {title} {Side-channel-free
  quantum key distribution},\ }\href@noop {} {\bibfield  {journal} {\bibinfo
  {journal} {Phys. Rev. Lett.}\ }\textbf {\bibinfo {volume} {108}},\ \bibinfo
  {pages} {130502} (\bibinfo {year} {2012})}\BibitemShut {NoStop}%
\bibitem [{\citenamefont {Lucamarini}\ \emph {et~al.}(2018)\citenamefont
  {Lucamarini}, \citenamefont {Yuan}, \citenamefont {Dynes},\ and\
  \citenamefont {Shields}}]{lucamarini2018overcoming}%
  \BibitemOpen
  \bibfield  {author} {\bibinfo {author} {\bibfnamefont {M.}~\bibnamefont
  {Lucamarini}}, \bibinfo {author} {\bibfnamefont {Z.~L.}\ \bibnamefont
  {Yuan}}, \bibinfo {author} {\bibfnamefont {J.~F.}\ \bibnamefont {Dynes}},\
  and\ \bibinfo {author} {\bibfnamefont {A.~J.}\ \bibnamefont {Shields}},\
  }\bibfield  {title} {\bibinfo {title} {Overcoming the rate--distance limit of
  quantum key distribution without quantum repeaters},\ }\href@noop {}
  {\bibfield  {journal} {\bibinfo  {journal} {Nature}\ }\textbf {\bibinfo
  {volume} {557}},\ \bibinfo {pages} {400} (\bibinfo {year}
  {2018})}\BibitemShut {NoStop}%
\bibitem [{\citenamefont {Ma}\ \emph {et~al.}(2018)\citenamefont {Ma},
  \citenamefont {Zeng},\ and\ \citenamefont {Zhou}}]{ma2018phase}%
  \BibitemOpen
  \bibfield  {author} {\bibinfo {author} {\bibfnamefont {X.}~\bibnamefont
  {Ma}}, \bibinfo {author} {\bibfnamefont {P.}~\bibnamefont {Zeng}},\ and\
  \bibinfo {author} {\bibfnamefont {H.}~\bibnamefont {Zhou}},\ }\bibfield
  {title} {\bibinfo {title} {Phase-matching quantum key distribution},\
  }\href@noop {} {\bibfield  {journal} {\bibinfo  {journal} {Physical Review
  X}\ }\textbf {\bibinfo {volume} {8}},\ \bibinfo {pages} {031043} (\bibinfo
  {year} {2018})}\BibitemShut {NoStop}%
\bibitem [{\citenamefont {Wang}\ \emph {et~al.}(2018)\citenamefont {Wang},
  \citenamefont {Yu},\ and\ \citenamefont {Hu}}]{wang2018twin}%
  \BibitemOpen
  \bibfield  {author} {\bibinfo {author} {\bibfnamefont {X.-B.}\ \bibnamefont
  {Wang}}, \bibinfo {author} {\bibfnamefont {Z.-W.}\ \bibnamefont {Yu}},\ and\
  \bibinfo {author} {\bibfnamefont {X.-L.}\ \bibnamefont {Hu}},\ }\bibfield
  {title} {\bibinfo {title} {Twin-field quantum key distribution with large
  misalignment error},\ }\href@noop {} {\bibfield  {journal} {\bibinfo
  {journal} {Physical Review A}\ }\textbf {\bibinfo {volume} {98}},\ \bibinfo
  {pages} {062323} (\bibinfo {year} {2018})}\BibitemShut {NoStop}%
\bibitem [{\citenamefont {Lin}\ and\ \citenamefont
  {L{\"u}tkenhaus}(2018)}]{lin2018simple}%
  \BibitemOpen
  \bibfield  {author} {\bibinfo {author} {\bibfnamefont {J.}~\bibnamefont
  {Lin}}\ and\ \bibinfo {author} {\bibfnamefont {N.}~\bibnamefont
  {L{\"u}tkenhaus}},\ }\bibfield  {title} {\bibinfo {title} {Simple security
  analysis of phase-matching measurement-device-independent quantum key
  distribution},\ }\href@noop {} {\bibfield  {journal} {\bibinfo  {journal}
  {Physical Review A}\ }\textbf {\bibinfo {volume} {98}},\ \bibinfo {pages}
  {042332} (\bibinfo {year} {2018})}\BibitemShut {NoStop}%
\bibitem [{\citenamefont {Curty}\ \emph {et~al.}(2019)\citenamefont {Curty},
  \citenamefont {Azuma},\ and\ \citenamefont {Lo}}]{curty2019simple}%
  \BibitemOpen
  \bibfield  {author} {\bibinfo {author} {\bibfnamefont {M.}~\bibnamefont
  {Curty}}, \bibinfo {author} {\bibfnamefont {K.}~\bibnamefont {Azuma}},\ and\
  \bibinfo {author} {\bibfnamefont {H.-K.}\ \bibnamefont {Lo}},\ }\bibfield
  {title} {\bibinfo {title} {Simple security proof of twin-field type quantum
  key distribution protocol},\ }\href@noop {} {\bibfield  {journal} {\bibinfo
  {journal} {npj Quantum Information}\ }\textbf {\bibinfo {volume} {5}},\
  \bibinfo {pages} {64} (\bibinfo {year} {2019})}\BibitemShut {NoStop}%
\bibitem [{\citenamefont {Cui}\ \emph {et~al.}(2019)\citenamefont {Cui},
  \citenamefont {Yin}, \citenamefont {Wang}, \citenamefont {Chen},
  \citenamefont {Wang}, \citenamefont {Guo},\ and\ \citenamefont
  {Han}}]{cui2019twin}%
  \BibitemOpen
  \bibfield  {author} {\bibinfo {author} {\bibfnamefont {C.}~\bibnamefont
  {Cui}}, \bibinfo {author} {\bibfnamefont {Z.-Q.}\ \bibnamefont {Yin}},
  \bibinfo {author} {\bibfnamefont {R.}~\bibnamefont {Wang}}, \bibinfo {author}
  {\bibfnamefont {W.}~\bibnamefont {Chen}}, \bibinfo {author} {\bibfnamefont
  {S.}~\bibnamefont {Wang}}, \bibinfo {author} {\bibfnamefont {G.-C.}\
  \bibnamefont {Guo}},\ and\ \bibinfo {author} {\bibfnamefont {Z.-F.}\
  \bibnamefont {Han}},\ }\bibfield  {title} {\bibinfo {title} {Twin-field
  quantum key distribution without phase postselection},\ }\href@noop {}
  {\bibfield  {journal} {\bibinfo  {journal} {Physical Review Applied}\
  }\textbf {\bibinfo {volume} {11}},\ \bibinfo {pages} {034053} (\bibinfo
  {year} {2019})}\BibitemShut {NoStop}%
\bibitem [{\citenamefont {Zeng}\ \emph {et~al.}(2022)\citenamefont {Zeng},
  \citenamefont {Zhou}, \citenamefont {Wu},\ and\ \citenamefont
  {Ma}}]{zeng2022mode}%
  \BibitemOpen
  \bibfield  {author} {\bibinfo {author} {\bibfnamefont {P.}~\bibnamefont
  {Zeng}}, \bibinfo {author} {\bibfnamefont {H.}~\bibnamefont {Zhou}}, \bibinfo
  {author} {\bibfnamefont {W.}~\bibnamefont {Wu}},\ and\ \bibinfo {author}
  {\bibfnamefont {X.}~\bibnamefont {Ma}},\ }\bibfield  {title} {\bibinfo
  {title} {Mode-pairing quantum key distribution},\ }\href@noop {} {\bibfield
  {journal} {\bibinfo  {journal} {Nature Communications}\ }\textbf {\bibinfo
  {volume} {13}},\ \bibinfo {pages} {3903} (\bibinfo {year}
  {2022})}\BibitemShut {NoStop}%
\bibitem [{\citenamefont {Xie}\ \emph {et~al.}(2022)\citenamefont {Xie},
  \citenamefont {Lu}, \citenamefont {Weng}, \citenamefont {Cao}, \citenamefont
  {Jia}, \citenamefont {Bao}, \citenamefont {Wang}, \citenamefont {Fu},
  \citenamefont {Yin},\ and\ \citenamefont {Chen}}]{xie2022breaking}%
  \BibitemOpen
  \bibfield  {author} {\bibinfo {author} {\bibfnamefont {Y.-M.}\ \bibnamefont
  {Xie}}, \bibinfo {author} {\bibfnamefont {Y.-S.}\ \bibnamefont {Lu}},
  \bibinfo {author} {\bibfnamefont {C.-X.}\ \bibnamefont {Weng}}, \bibinfo
  {author} {\bibfnamefont {X.-Y.}\ \bibnamefont {Cao}}, \bibinfo {author}
  {\bibfnamefont {Z.-Y.}\ \bibnamefont {Jia}}, \bibinfo {author} {\bibfnamefont
  {Y.}~\bibnamefont {Bao}}, \bibinfo {author} {\bibfnamefont {Y.}~\bibnamefont
  {Wang}}, \bibinfo {author} {\bibfnamefont {Y.}~\bibnamefont {Fu}}, \bibinfo
  {author} {\bibfnamefont {H.-L.}\ \bibnamefont {Yin}},\ and\ \bibinfo {author}
  {\bibfnamefont {Z.-B.}\ \bibnamefont {Chen}},\ }\bibfield  {title} {\bibinfo
  {title} {Breaking the rate-loss bound of quantum key distribution with
  asynchronous two-photon interference},\ }\href@noop {} {\bibfield  {journal}
  {\bibinfo  {journal} {PRX Quantum}\ }\textbf {\bibinfo {volume} {3}},\
  \bibinfo {pages} {020315} (\bibinfo {year} {2022})}\BibitemShut {NoStop}%
\bibitem [{\citenamefont {Pirandola}\ \emph {et~al.}(2009)\citenamefont
  {Pirandola}, \citenamefont {Garc{\'\i}a-Patr{\'o}n}, \citenamefont
  {Braunstein},\ and\ \citenamefont {Lloyd}}]{pirandola2009direct}%
  \BibitemOpen
  \bibfield  {author} {\bibinfo {author} {\bibfnamefont {S.}~\bibnamefont
  {Pirandola}}, \bibinfo {author} {\bibfnamefont {R.}~\bibnamefont
  {Garc{\'\i}a-Patr{\'o}n}}, \bibinfo {author} {\bibfnamefont {S.~L.}\
  \bibnamefont {Braunstein}},\ and\ \bibinfo {author} {\bibfnamefont
  {S.}~\bibnamefont {Lloyd}},\ }\bibfield  {title} {\bibinfo {title} {Direct
  and reverse secret-key capacities of a quantum channel},\ }\href@noop {}
  {\bibfield  {journal} {\bibinfo  {journal} {Physical review letters}\
  }\textbf {\bibinfo {volume} {102}},\ \bibinfo {pages} {050503} (\bibinfo
  {year} {2009})}\BibitemShut {NoStop}%
\bibitem [{\citenamefont {Takeoka}\ \emph {et~al.}(2014)\citenamefont
  {Takeoka}, \citenamefont {Guha},\ and\ \citenamefont
  {Wilde}}]{takeoka2014fundamental}%
  \BibitemOpen
  \bibfield  {author} {\bibinfo {author} {\bibfnamefont {M.}~\bibnamefont
  {Takeoka}}, \bibinfo {author} {\bibfnamefont {S.}~\bibnamefont {Guha}},\ and\
  \bibinfo {author} {\bibfnamefont {M.~M.}\ \bibnamefont {Wilde}},\ }\bibfield
  {title} {\bibinfo {title} {Fundamental rate-loss tradeoff for optical quantum
  key distribution},\ }\href@noop {} {\bibfield  {journal} {\bibinfo  {journal}
  {Nature communications}\ }\textbf {\bibinfo {volume} {5}},\ \bibinfo {pages}
  {5235} (\bibinfo {year} {2014})}\BibitemShut {NoStop}%
\bibitem [{\citenamefont {Pirandola}\ \emph {et~al.}(2017)\citenamefont
  {Pirandola}, \citenamefont {Laurenza}, \citenamefont {Ottaviani},\ and\
  \citenamefont {Banchi}}]{pirandola2017fundamental}%
  \BibitemOpen
  \bibfield  {author} {\bibinfo {author} {\bibfnamefont {S.}~\bibnamefont
  {Pirandola}}, \bibinfo {author} {\bibfnamefont {R.}~\bibnamefont {Laurenza}},
  \bibinfo {author} {\bibfnamefont {C.}~\bibnamefont {Ottaviani}},\ and\
  \bibinfo {author} {\bibfnamefont {L.}~\bibnamefont {Banchi}},\ }\bibfield
  {title} {\bibinfo {title} {Fundamental limits of repeaterless quantum
  communications},\ }\href@noop {} {\bibfield  {journal} {\bibinfo  {journal}
  {Nature communications}\ }\textbf {\bibinfo {volume} {8}},\ \bibinfo {pages}
  {15043} (\bibinfo {year} {2017})}\BibitemShut {NoStop}%
\bibitem [{\citenamefont {Tamaki}\ \emph {et~al.}(2014)\citenamefont {Tamaki},
  \citenamefont {Curty}, \citenamefont {Kato}, \citenamefont {Lo},\ and\
  \citenamefont {Azuma}}]{tamaki2014loss}%
  \BibitemOpen
  \bibfield  {author} {\bibinfo {author} {\bibfnamefont {K.}~\bibnamefont
  {Tamaki}}, \bibinfo {author} {\bibfnamefont {M.}~\bibnamefont {Curty}},
  \bibinfo {author} {\bibfnamefont {G.}~\bibnamefont {Kato}}, \bibinfo {author}
  {\bibfnamefont {H.-K.}\ \bibnamefont {Lo}},\ and\ \bibinfo {author}
  {\bibfnamefont {K.}~\bibnamefont {Azuma}},\ }\bibfield  {title} {\bibinfo
  {title} {Loss-tolerant quantum cryptography with imperfect sources},\
  }\href@noop {} {\bibfield  {journal} {\bibinfo  {journal} {Physical Review
  A}\ }\textbf {\bibinfo {volume} {90}},\ \bibinfo {pages} {052314} (\bibinfo
  {year} {2014})}\BibitemShut {NoStop}%
\bibitem [{\citenamefont {Pereira}\ \emph {et~al.}(2019)\citenamefont
  {Pereira}, \citenamefont {Curty},\ and\ \citenamefont
  {Tamaki}}]{pereira2019quantum}%
  \BibitemOpen
  \bibfield  {author} {\bibinfo {author} {\bibfnamefont {M.}~\bibnamefont
  {Pereira}}, \bibinfo {author} {\bibfnamefont {M.}~\bibnamefont {Curty}},\
  and\ \bibinfo {author} {\bibfnamefont {K.}~\bibnamefont {Tamaki}},\
  }\bibfield  {title} {\bibinfo {title} {Quantum key distribution with flawed
  and leaky sources},\ }\href@noop {} {\bibfield  {journal} {\bibinfo
  {journal} {npj Quantum Information}\ }\textbf {\bibinfo {volume} {5}},\
  \bibinfo {pages} {62} (\bibinfo {year} {2019})}\BibitemShut {NoStop}%
\bibitem [{\citenamefont {Wang}\ \emph {et~al.}(2019)\citenamefont {Wang},
  \citenamefont {Hu},\ and\ \citenamefont {Yu}}]{wang2019practical}%
  \BibitemOpen
  \bibfield  {author} {\bibinfo {author} {\bibfnamefont {X.-B.}\ \bibnamefont
  {Wang}}, \bibinfo {author} {\bibfnamefont {X.-L.}\ \bibnamefont {Hu}},\ and\
  \bibinfo {author} {\bibfnamefont {Z.-W.}\ \bibnamefont {Yu}},\ }\bibfield
  {title} {\bibinfo {title} {Practical long-distance side-channel-free quantum
  key distribution},\ }\href@noop {} {\bibfield  {journal} {\bibinfo  {journal}
  {Physical Review Applied}\ }\textbf {\bibinfo {volume} {12}},\ \bibinfo
  {pages} {054034} (\bibinfo {year} {2019})}\BibitemShut {NoStop}%
\bibitem [{\citenamefont {Pereira}\ \emph {et~al.}(2020)\citenamefont
  {Pereira}, \citenamefont {Kato}, \citenamefont {Mizutani}, \citenamefont
  {Curty},\ and\ \citenamefont {Tamaki}}]{pereira2020quantum}%
  \BibitemOpen
  \bibfield  {author} {\bibinfo {author} {\bibfnamefont {M.}~\bibnamefont
  {Pereira}}, \bibinfo {author} {\bibfnamefont {G.}~\bibnamefont {Kato}},
  \bibinfo {author} {\bibfnamefont {A.}~\bibnamefont {Mizutani}}, \bibinfo
  {author} {\bibfnamefont {M.}~\bibnamefont {Curty}},\ and\ \bibinfo {author}
  {\bibfnamefont {K.}~\bibnamefont {Tamaki}},\ }\bibfield  {title} {\bibinfo
  {title} {Quantum key distribution with correlated sources},\ }\href@noop {}
  {\bibfield  {journal} {\bibinfo  {journal} {Science Advances}\ }\textbf
  {\bibinfo {volume} {6}},\ \bibinfo {pages} {eaaz4487} (\bibinfo {year}
  {2020})}\BibitemShut {NoStop}%
\bibitem [{\citenamefont {Navarrete}\ \emph {et~al.}(2021)\citenamefont
  {Navarrete}, \citenamefont {Pereira}, \citenamefont {Curty},\ and\
  \citenamefont {Tamaki}}]{navarrete2021practical}%
  \BibitemOpen
  \bibfield  {author} {\bibinfo {author} {\bibfnamefont {{\'A}.}~\bibnamefont
  {Navarrete}}, \bibinfo {author} {\bibfnamefont {M.}~\bibnamefont {Pereira}},
  \bibinfo {author} {\bibfnamefont {M.}~\bibnamefont {Curty}},\ and\ \bibinfo
  {author} {\bibfnamefont {K.}~\bibnamefont {Tamaki}},\ }\bibfield  {title}
  {\bibinfo {title} {Practical quantum key distribution that is secure against
  side channels},\ }\href@noop {} {\bibfield  {journal} {\bibinfo  {journal}
  {Physical Review Applied}\ }\textbf {\bibinfo {volume} {15}},\ \bibinfo
  {pages} {034072} (\bibinfo {year} {2021})}\BibitemShut {NoStop}%
\bibitem [{\citenamefont {Gu}\ \emph {et~al.}(2022)\citenamefont {Gu},
  \citenamefont {Cao}, \citenamefont {Fu}, \citenamefont {He}, \citenamefont
  {Yin}, \citenamefont {Yin},\ and\ \citenamefont {Chen}}]{gu2022experimental}%
  \BibitemOpen
  \bibfield  {author} {\bibinfo {author} {\bibfnamefont {J.}~\bibnamefont
  {Gu}}, \bibinfo {author} {\bibfnamefont {X.-Y.}\ \bibnamefont {Cao}},
  \bibinfo {author} {\bibfnamefont {Y.}~\bibnamefont {Fu}}, \bibinfo {author}
  {\bibfnamefont {Z.-W.}\ \bibnamefont {He}}, \bibinfo {author} {\bibfnamefont
  {Z.-J.}\ \bibnamefont {Yin}}, \bibinfo {author} {\bibfnamefont {H.-L.}\
  \bibnamefont {Yin}},\ and\ \bibinfo {author} {\bibfnamefont {Z.-B.}\
  \bibnamefont {Chen}},\ }\bibfield  {title} {\bibinfo {title} {Experimental
  measurement-device-independent type quantum key distribution with flawed and
  correlated sources},\ }\href@noop {} {\bibfield  {journal} {\bibinfo
  {journal} {Science Bulletin}\ }\textbf {\bibinfo {volume} {67}},\ \bibinfo
  {pages} {2167} (\bibinfo {year} {2022})}\BibitemShut {NoStop}%
\bibitem [{\citenamefont {Zhang}\ \emph {et~al.}(2022)\citenamefont {Zhang},
  \citenamefont {Hu}, \citenamefont {Jiang}, \citenamefont {Chen},
  \citenamefont {Liu}, \citenamefont {Zhang}, \citenamefont {Yu}, \citenamefont
  {Li}, \citenamefont {You}, \citenamefont {Wang} \emph
  {et~al.}}]{zhang2022experimental}%
  \BibitemOpen
  \bibfield  {author} {\bibinfo {author} {\bibfnamefont {C.}~\bibnamefont
  {Zhang}}, \bibinfo {author} {\bibfnamefont {X.-L.}\ \bibnamefont {Hu}},
  \bibinfo {author} {\bibfnamefont {C.}~\bibnamefont {Jiang}}, \bibinfo
  {author} {\bibfnamefont {J.-P.}\ \bibnamefont {Chen}}, \bibinfo {author}
  {\bibfnamefont {Y.}~\bibnamefont {Liu}}, \bibinfo {author} {\bibfnamefont
  {W.}~\bibnamefont {Zhang}}, \bibinfo {author} {\bibfnamefont {Z.-W.}\
  \bibnamefont {Yu}}, \bibinfo {author} {\bibfnamefont {H.}~\bibnamefont {Li}},
  \bibinfo {author} {\bibfnamefont {L.}~\bibnamefont {You}}, \bibinfo {author}
  {\bibfnamefont {Z.}~\bibnamefont {Wang}}, \emph {et~al.},\ }\bibfield
  {title} {\bibinfo {title} {Experimental side-channel-secure quantum key
  distribution},\ }\href@noop {} {\bibfield  {journal} {\bibinfo  {journal}
  {Physical Review Letters}\ }\textbf {\bibinfo {volume} {128}},\ \bibinfo
  {pages} {190503} (\bibinfo {year} {2022})}\BibitemShut {NoStop}%
\bibitem [{\citenamefont {Jiang}\ \emph {et~al.}(2022)\citenamefont {Jiang},
  \citenamefont {Yu}, \citenamefont {Hu},\ and\ \citenamefont
  {Wang}}]{jiang2022side}%
  \BibitemOpen
  \bibfield  {author} {\bibinfo {author} {\bibfnamefont {C.}~\bibnamefont
  {Jiang}}, \bibinfo {author} {\bibfnamefont {Z.-W.}\ \bibnamefont {Yu}},
  \bibinfo {author} {\bibfnamefont {X.-L.}\ \bibnamefont {Hu}},\ and\ \bibinfo
  {author} {\bibfnamefont {X.-B.}\ \bibnamefont {Wang}},\ }\bibfield  {title}
  {\bibinfo {title} {Side-channel-free quantum key distribution with practical
  devices},\ }\href@noop {} {\bibfield  {journal} {\bibinfo  {journal} {arXiv
  preprint arXiv:2205.08421}\ } (\bibinfo {year} {2022})}\BibitemShut {NoStop}%
\bibitem [{\citenamefont {Hwang}(2003)}]{hwang2003quantum}%
  \BibitemOpen
  \bibfield  {author} {\bibinfo {author} {\bibfnamefont {W.-Y.}\ \bibnamefont
  {Hwang}},\ }\bibfield  {title} {\bibinfo {title} {Quantum key distribution
  with high loss: toward global secure communication},\ }\href@noop {}
  {\bibfield  {journal} {\bibinfo  {journal} {Physical review letters}\
  }\textbf {\bibinfo {volume} {91}},\ \bibinfo {pages} {057901} (\bibinfo
  {year} {2003})}\BibitemShut {NoStop}%
\bibitem [{\citenamefont {Lo}\ \emph {et~al.}(2005)\citenamefont {Lo},
  \citenamefont {Ma},\ and\ \citenamefont {Chen}}]{PhysRevLett.94.230504}%
  \BibitemOpen
  \bibfield  {author} {\bibinfo {author} {\bibfnamefont {H.-K.}\ \bibnamefont
  {Lo}}, \bibinfo {author} {\bibfnamefont {X.}~\bibnamefont {Ma}},\ and\
  \bibinfo {author} {\bibfnamefont {K.}~\bibnamefont {Chen}},\ }\bibfield
  {title} {\bibinfo {title} {Decoy state quantum key distribution},\ }\href
  {https://doi.org/10.1103/PhysRevLett.94.230504} {\bibfield  {journal}
  {\bibinfo  {journal} {Phys. Rev. Lett.}\ }\textbf {\bibinfo {volume} {94}},\
  \bibinfo {pages} {230504} (\bibinfo {year} {2005})}\BibitemShut {NoStop}%
\bibitem [{\citenamefont {Wang}(2005)}]{wang2005beating}%
  \BibitemOpen
  \bibfield  {author} {\bibinfo {author} {\bibfnamefont {X.-B.}\ \bibnamefont
  {Wang}},\ }\bibfield  {title} {\bibinfo {title} {Beating the
  photon-number-splitting attack in practical quantum cryptography},\
  }\href@noop {} {\bibfield  {journal} {\bibinfo  {journal} {Physical review
  letters}\ }\textbf {\bibinfo {volume} {94}},\ \bibinfo {pages} {230503}
  (\bibinfo {year} {2005})}\BibitemShut {NoStop}%
\bibitem [{\citenamefont {Ma}\ \emph {et~al.}(2005)\citenamefont {Ma},
  \citenamefont {Qi}, \citenamefont {Zhao},\ and\ \citenamefont
  {Lo}}]{ma2005practical}%
  \BibitemOpen
  \bibfield  {author} {\bibinfo {author} {\bibfnamefont {X.}~\bibnamefont
  {Ma}}, \bibinfo {author} {\bibfnamefont {B.}~\bibnamefont {Qi}}, \bibinfo
  {author} {\bibfnamefont {Y.}~\bibnamefont {Zhao}},\ and\ \bibinfo {author}
  {\bibfnamefont {H.-K.}\ \bibnamefont {Lo}},\ }\bibfield  {title} {\bibinfo
  {title} {Practical decoy state for quantum key distribution},\ }\href@noop {}
  {\bibfield  {journal} {\bibinfo  {journal} {Physical Review A}\ }\textbf
  {\bibinfo {volume} {72}},\ \bibinfo {pages} {012326} (\bibinfo {year}
  {2005})}\BibitemShut {NoStop}%
\bibitem [{\citenamefont {Kato}(2020)}]{kato2020concentration}%
  \BibitemOpen
  \bibfield  {author} {\bibinfo {author} {\bibfnamefont {G.}~\bibnamefont
  {Kato}},\ }\bibfield  {title} {\bibinfo {title} {Concentration inequality
  using unconfirmed knowledge},\ }\href@noop {} {\bibfield  {journal} {\bibinfo
   {journal} {arXiv preprint arXiv:2002.04357}\ } (\bibinfo {year}
  {2020})}\BibitemShut {NoStop}%
\bibitem [{\citenamefont {Mitzenmacher}\ and\ \citenamefont
  {Upfal}(2017)}]{mitzenmacher2017probability}%
  \BibitemOpen
  \bibfield  {author} {\bibinfo {author} {\bibfnamefont {M.}~\bibnamefont
  {Mitzenmacher}}\ and\ \bibinfo {author} {\bibfnamefont {E.}~\bibnamefont
  {Upfal}},\ }\href@noop {} {\emph {\bibinfo {title} {Probability and
  computing: Randomization and probabilistic techniques in algorithms and data
  analysis}}}\ (\bibinfo  {publisher} {Cambridge university press},\ \bibinfo
  {year} {2017})\BibitemShut {NoStop}%
\bibitem [{\citenamefont {M{\"u}ller-Quade}\ and\ \citenamefont
  {Renner}(2009)}]{muller2009composability}%
  \BibitemOpen
  \bibfield  {author} {\bibinfo {author} {\bibfnamefont {J.}~\bibnamefont
  {M{\"u}ller-Quade}}\ and\ \bibinfo {author} {\bibfnamefont {R.}~\bibnamefont
  {Renner}},\ }\bibfield  {title} {\bibinfo {title} {Composability in quantum
  cryptography},\ }\href@noop {} {\bibfield  {journal} {\bibinfo  {journal}
  {New Journal of Physics}\ }\textbf {\bibinfo {volume} {11}},\ \bibinfo
  {pages} {085006} (\bibinfo {year} {2009})}\BibitemShut {NoStop}%
\bibitem [{\citenamefont {Tomamichel}\ \emph {et~al.}(2012)\citenamefont
  {Tomamichel}, \citenamefont {Lim}, \citenamefont {Gisin},\ and\ \citenamefont
  {Renner}}]{tomamichel2012tight}%
  \BibitemOpen
  \bibfield  {author} {\bibinfo {author} {\bibfnamefont {M.}~\bibnamefont
  {Tomamichel}}, \bibinfo {author} {\bibfnamefont {C.~C.~W.}\ \bibnamefont
  {Lim}}, \bibinfo {author} {\bibfnamefont {N.}~\bibnamefont {Gisin}},\ and\
  \bibinfo {author} {\bibfnamefont {R.}~\bibnamefont {Renner}},\ }\bibfield
  {title} {\bibinfo {title} {Tight finite-key analysis for quantum
  cryptography},\ }\href@noop {} {\bibfield  {journal} {\bibinfo  {journal}
  {Nature communications}\ }\textbf {\bibinfo {volume} {3}},\ \bibinfo {pages}
  {634} (\bibinfo {year} {2012})}\BibitemShut {NoStop}%
\bibitem [{\citenamefont {Yoshino}\ \emph {et~al.}(2018)\citenamefont
  {Yoshino}, \citenamefont {Fujiwara}, \citenamefont {Nakata}, \citenamefont
  {Sumiya}, \citenamefont {Sasaki}, \citenamefont {Takeoka}, \citenamefont
  {Sasaki}, \citenamefont {Tajima}, \citenamefont {Koashi},\ and\ \citenamefont
  {Tomita}}]{yoshino2018quantum}%
  \BibitemOpen
  \bibfield  {author} {\bibinfo {author} {\bibfnamefont {K.-i.}\ \bibnamefont
  {Yoshino}}, \bibinfo {author} {\bibfnamefont {M.}~\bibnamefont {Fujiwara}},
  \bibinfo {author} {\bibfnamefont {K.}~\bibnamefont {Nakata}}, \bibinfo
  {author} {\bibfnamefont {T.}~\bibnamefont {Sumiya}}, \bibinfo {author}
  {\bibfnamefont {T.}~\bibnamefont {Sasaki}}, \bibinfo {author} {\bibfnamefont
  {M.}~\bibnamefont {Takeoka}}, \bibinfo {author} {\bibfnamefont
  {M.}~\bibnamefont {Sasaki}}, \bibinfo {author} {\bibfnamefont
  {A.}~\bibnamefont {Tajima}}, \bibinfo {author} {\bibfnamefont
  {M.}~\bibnamefont {Koashi}},\ and\ \bibinfo {author} {\bibfnamefont
  {A.}~\bibnamefont {Tomita}},\ }\bibfield  {title} {\bibinfo {title} {Quantum
  key distribution with an efficient countermeasure against correlated
  intensity fluctuations in optical pulses},\ }\href@noop {} {\bibfield
  {journal} {\bibinfo  {journal} {npj Quantum Information}\ }\textbf {\bibinfo
  {volume} {4}},\ \bibinfo {pages} {8} (\bibinfo {year} {2018})}\BibitemShut
  {NoStop}%
\bibitem [{\citenamefont {Kang}\ \emph {et~al.}(2023)\citenamefont {Kang},
  \citenamefont {Lu}, \citenamefont {Wang}, \citenamefont {Chen}, \citenamefont
  {Wang}, \citenamefont {Yin}, \citenamefont {He}, \citenamefont {Chen},
  \citenamefont {Fan-Yuan}, \citenamefont {Guo} \emph
  {et~al.}}]{kang2023patterning}%
  \BibitemOpen
  \bibfield  {author} {\bibinfo {author} {\bibfnamefont {X.}~\bibnamefont
  {Kang}}, \bibinfo {author} {\bibfnamefont {F.-Y.}\ \bibnamefont {Lu}},
  \bibinfo {author} {\bibfnamefont {S.}~\bibnamefont {Wang}}, \bibinfo {author}
  {\bibfnamefont {J.-L.}\ \bibnamefont {Chen}}, \bibinfo {author}
  {\bibfnamefont {Z.-H.}\ \bibnamefont {Wang}}, \bibinfo {author}
  {\bibfnamefont {Z.-Q.}\ \bibnamefont {Yin}}, \bibinfo {author} {\bibfnamefont
  {D.-Y.}\ \bibnamefont {He}}, \bibinfo {author} {\bibfnamefont
  {W.}~\bibnamefont {Chen}}, \bibinfo {author} {\bibfnamefont {G.-J.}\
  \bibnamefont {Fan-Yuan}}, \bibinfo {author} {\bibfnamefont {G.-C.}\
  \bibnamefont {Guo}}, \emph {et~al.},\ }\bibfield  {title} {\bibinfo {title}
  {Patterning-effect calibration algorithm for secure decoy-state quantum key
  distribution},\ }\href@noop {} {\bibfield  {journal} {\bibinfo  {journal}
  {Journal of Lightwave Technology}\ }\textbf {\bibinfo {volume} {41}},\
  \bibinfo {pages} {75} (\bibinfo {year} {2023})}\BibitemShut {NoStop}%
\bibitem [{\citenamefont {Jiang}\ \emph {et~al.}(2023)\citenamefont {Jiang},
  \citenamefont {Hu}, \citenamefont {Yu},\ and\ \citenamefont
  {Wang}}]{jiang2023side}%
  \BibitemOpen
  \bibfield  {author} {\bibinfo {author} {\bibfnamefont {C.}~\bibnamefont
  {Jiang}}, \bibinfo {author} {\bibfnamefont {X.-L.}\ \bibnamefont {Hu}},
  \bibinfo {author} {\bibfnamefont {Z.-W.}\ \bibnamefont {Yu}},\ and\ \bibinfo
  {author} {\bibfnamefont {X.-B.}\ \bibnamefont {Wang}},\ }\bibfield  {title}
  {\bibinfo {title} {Side-channel-secure quantum key distribution},\
  }\href@noop {} {\bibfield  {journal} {\bibinfo  {journal} {arXiv preprint
  arXiv:2305.08148}\ } (\bibinfo {year} {2023})}\BibitemShut {NoStop}%
\bibitem [{\citenamefont {Azuma}(1967)}]{azuma1967weighted}%
  \BibitemOpen
  \bibfield  {author} {\bibinfo {author} {\bibfnamefont {K.}~\bibnamefont
  {Azuma}},\ }\bibfield  {title} {\bibinfo {title} {Weighted sums of certain
  dependent random variables},\ }\href@noop {} {\bibfield  {journal} {\bibinfo
  {journal} {Tohoku Mathematical Journal, Second Series}\ }\textbf {\bibinfo
  {volume} {19}},\ \bibinfo {pages} {357} (\bibinfo {year} {1967})}\BibitemShut
  {NoStop}%
\bibitem [{\citenamefont {Curr{\'a}s-Lorenzo}\ \emph
  {et~al.}(2021)\citenamefont {Curr{\'a}s-Lorenzo}, \citenamefont {Navarrete},
  \citenamefont {Azuma}, \citenamefont {Kato}, \citenamefont {Curty},\ and\
  \citenamefont {Razavi}}]{curras2021tight}%
  \BibitemOpen
  \bibfield  {author} {\bibinfo {author} {\bibfnamefont {G.}~\bibnamefont
  {Curr{\'a}s-Lorenzo}}, \bibinfo {author} {\bibfnamefont {{\'A}.}~\bibnamefont
  {Navarrete}}, \bibinfo {author} {\bibfnamefont {K.}~\bibnamefont {Azuma}},
  \bibinfo {author} {\bibfnamefont {G.}~\bibnamefont {Kato}}, \bibinfo {author}
  {\bibfnamefont {M.}~\bibnamefont {Curty}},\ and\ \bibinfo {author}
  {\bibfnamefont {M.}~\bibnamefont {Razavi}},\ }\bibfield  {title} {\bibinfo
  {title} {Tight finite-key security for twin-field quantum key distribution},\
  }\href@noop {} {\bibfield  {journal} {\bibinfo  {journal} {npj Quantum
  Information}\ }\textbf {\bibinfo {volume} {7}},\ \bibinfo {pages} {22}
  (\bibinfo {year} {2021})}\BibitemShut {NoStop}%
\end{thebibliography}%
\end{document}